\documentclass{article}
\usepackage[latin1]{inputenc}
\usepackage[english]{babel}
\usepackage[dvips]{graphics}
\usepackage{epsfig}
\usepackage{fullpage}
\usepackage{amsfonts}
\usepackage{multicol}
\usepackage{latexsym,amssymb}     
\usepackage{amsmath}
\usepackage{indentfirst}
\usepackage{makeidx}
\usepackage{theorem}
\usepackage{bbold}
\usepackage{psfrag}

\newcommand{\rr}{\mathbb{R}}
\newcommand{\ee}{\mathbb{E}}
\newcommand{\qq}{\mathbb{Q}}
\newcommand{\nn}{\mathbb{N}}
\newcommand{\pp}{\mathbb{P}}
\newcommand{\ov}{\operatornamewithlimits}
\newcommand{\ds}{\displaystyle}
\newcommand{\ind}{\mathbb{1}}

\def\BS{Black \& Scholes}

\newcommand{\blanc}[1]{\vspace{#1\baselineskip}}

\newtheorem{nt_theorem}{Theorem}

\newtheorem{nt_proposition}[nt_theorem]{Proposition}
\newenvironment{proposition}{\blanc{1.5}\begin{nt_proposition}---}{\end{nt_proposition}\blanc{1.5}}

\newtheorem{nt_definition}[nt_theorem]{Definition}

\newtheorem{nt_lemma}[nt_theorem]{Lemma}
\newenvironment{lemma}{\blanc{0.5}\begin{nt_lemma}---}{\end{nt_lemma}\blanc{0.5}}

\newtheorem{nt_jeulemma}[nt_theorem]{Jeulin's Lemma}

\newtheorem{nt_conjecture}[nt_theorem]{Conjecture}

\newtheorem{nt_remark}[nt_theorem]{Remark}
\newenvironment{remark}{\blanc{0.2}\begin{nt_remark}---}{\end{nt_remark}\blanc{0.2}}

\newtheorem{algo}{Algorithm}

\newenvironment{proof}{{\textit{Proof : }}}{\hfill$\Box$
\\\bigskip}

\begin{document}

\vspace{5mm}

\begin{center}
\textbf{\Large{Exact retrospective Monte Carlo computation\\ of arithmetic average Asian options}}\\
\end{center}

$\,$

\begin{center}
\emph{Benjamin Jourdain}\footnote{Project team Math Fi, CERMICS, Ecole des Ponts, Paristech, supported by the ANR program ADAP'MC.

Postal address : 6-8 av. Blaise Pascal, Cité Descartes,
Champs-sur-Marne, 77455 Marne-la-Vallée Cedex 2.

E-mails : jourdain@cermics.enpc.fr and sbai@cermics.enpc.fr
} and \emph{Mohamed Sbai$\,^1$}
\end{center}

\vspace{20mm}

\begin{abstract}
Taking advantage of the recent literature on exact simulation
algorithms (Beskos \textit{et al.} \cite{BPR}) and unbiased
estimation of the expectation of certain functional integrals
(Wagner \cite{Wagner3_1}, Beskos \textit{et al.} \cite{BPRF} and
Fearnhead \textit{et al.} \cite{FPR}), we apply an exact simulation
based technique for pricing continuous arithmetic average Asian
options in the \BS~framework. Unlike existing Monte Carlo methods,
we are no longer prone to the discretization bias resulting from the
approximation of continuous time processes through discrete
sampling. Numerical results of simulation studies are presented and
variance reduction problems are considered.
\end{abstract}

\vspace{15mm}

\section*{Introduction}
Although the \BS~framework is very simple, it is still a challenging
task to efficiently price Asian options. Since we do not know
explicitly the distribution of the arithmetic sum of log-normal
variables, there is no closed form solution for the price of an
Asian option. By the early nineties, many researchers attempted to
address this problem and hence different approaches were studied
including analytic approximations (see Turnball and Wakeman
\cite{TurnballWakeman}, Vorst \cite{Vorst}, Levy \cite{Levy} and
more recently Lord \cite{Lord}), PDE methods (see Vecer
\cite{Vecer}, Rogers and Shi \cite{RogersShi}, Ingersoll
\cite{Ingersoll}, Dubois and Lelievre \cite{Lelievre}), Laplace
transform inversion methods (see Geman and Yor \cite{GemanYor},
Geman and Eydeland \cite{GemanEydeland}) and, of course, Monte Carlo
simulation methods (see Kemna and Vorst \cite{KemnaVorst}, Broadie
and Glasserman \cite{BroadieGlasserman}, Fu \textit{et al.}
\cite{FuMadanWang}).

Monte Carlo simulation can be computationally expensive because of
the usual statistical error. Variance reduction techniques are then
essential to accelerate the convergence (one of the most efficient
techniques is the Kemna\&Vorst control variate based on the
geometric average). One must also account for the inherent
discretization bias resulting from approximating the continuous
average of the stock price with a discrete one. It is crucial to
choose with care the discretization scheme in order to have an
accurate solution (see Lapeyre and Temam \cite{LapeyreTemam}). The
main contribution of our work is to fully address this last feature
by the use, after a suitable change of variables, of an exact
simulation method inspired from the recent work of Beskos \textit{et
al.} \cite{BPR,BPRF} and Fearnhead \textit{et al.} \cite{FPR}.

In the first part of the paper, we recall the algorithm introduced
by Beskos \textit{et al.} \cite{BPR} in order to simulate
sample-paths of processes solving one-dimensional stochastic
differential equations. By a suitable change of variables, one may
suppose that the diffusion coefficient is equal to one. Then,
according to the Girsanov theorem, one may deal with the drift
coefficient by introducing an exponential martingale weight. Because
of the one-dimensional setting, the stochastic integral in this
exponential weight is equal to a standard integral with respect to
the time variable up to the addition of a function of the terminal
value of the path. Under suitable assumptions, conditionally on a
Brownian path, an event with probability equal to the normalized
exponential weight can be simulated using a Poisson point process.
This allows to accept or reject this Brownian path as a path
solution to the SDE with diffusion coefficient equal to one. In
finance, one is interested in computing expectations rather than
exact simulation of the paths. In this perspective, computation of
the exponential importance sampling weight is enough. The entire
series expansion of the exponential function permits to replace this
exponential weight by a computable weight with the same conditional
expectation given the Brownian path. This idea was first introduced
by Wagner \cite{Wagner3_1,Wagner2,Wagner3_2,Wagner1} in a
statistical physics context and it was very recently revisited by
Beskos \textit{et al.} \cite{BPRF} and Fearnhead \textit{et al.}
\cite{FPR} for the estimation of partially observed diffusions. Some
of the assumptions necessary to implement the exact algorithm of
Beskos \textit{et al.} \cite{BPR} can then be weakened.

The second part is devoted to the application of
these methods to option pricing within the\\
\BS~framework. Throughout the paper, $\ds S_t=S_0\exp\left(\sigma
W_t+(r-\delta-\frac{\sigma^2}{2})t\right)$ represents the stock
price at time $t$, $T$ the maturity of the option, $r$ the short
interest rate, $\sigma$ the volatility parameter, $\delta$ the
dividend rate and $(W)_{t\in [0,T]}$ denotes a standard Brownian
motion on the risk-neutral probability space
$(\Omega,\mathcal{F},\pp)$. We are interested in computing the price
$C_0=\ee\left(e^{-rT}f\left(\alpha S_T + \beta
\int_0^TS_tdt\right)\right)$ of a European option with pay-off
$f\left(\alpha S_T + \beta \int_0^TS_tdt\right)$ assumed to be
square integrable under the risk neutral measure $\pp$. The
constants $\alpha$ and $\beta$ are two given non-negative
parameters.

When $\alpha>0$, we remark that, by a change of variables inspired
by Rogers and Shi \cite{RogersShi}, $\alpha S_T+\beta\int_0^TS_tdt$
has the same law as the solution at time $T$ of a well-chosen
one-dimensional stochastic differential equation. Then it is easy to
implement the exact methods previously presented. The case
$\alpha=0$ of standard Asian options is more intricate. The previous
approach does not work and we propose a new change of variables
which is singular at initial time. It is not possible to implement
neither the exact simulation algorithm nor the method based on the
unbiased estimator of Wagner \cite{Wagner3_1} and we propose a
pseudo-exact hybrid method which appears as an extension of the
exact simulation algorithm. In both cases, one first replaces the
integral with respect to the time variable in the function $f$ by an
integral with respect to time in the exponential function. Because
of the nice properties of this last function, exact computation is
possible.


\section{Exact Simulation techniques}
\subsection{The exact simulation method of Beskos \textit{et
al.} \cite{BPR}} In a recent paper, Beskos \textit{et al.}
\cite{BPR} proposed an algorithm which allows to simulate exactly
the solution of a 1-dimensional stochastic differential equation.
Under some hypotheses, they manage to implement an
acceptance-rejection algorithm over the whole path of the solution,
based on recursive simulation of a biased Brownian motion. Let us
briefly recall their methodology. We refer to~\cite{BPR} for the
demonstrations and a detailed presentation.

Consider the stochastic process $(\xi_t)_{0 \leq t \leq T}$
determined as the solution of a general stochastic differential
equation of the form :
\begin{equation}
\left\{\begin{array}{rcl}
d\xi_t&=&b(\xi_t)dt+\sigma(\xi_t) dW_t\\
\xi_0&=&\xi\in \rr
\end{array}\right.
\label{eqn:SDEo}
\end{equation}
where $b$ and $\sigma$ are scalar functions satisfying the usual
Lipschitz and growth conditions with $\sigma$ non vanishing. To
simplify this equation, Beskos \textit{et al.} \cite{BPR} suggest to
use the following change of variables : $X_t=\eta(\xi_t)$ where
$\eta$ is a primitive of $\frac{1}{\sigma}$ ($\eta(x)=\int_.^x
\frac{1}{\sigma(u)}
du$).\\
Under the additional assumption that $\frac{1}{\sigma}$ is
continuously differentiable, one can apply Itô's lemma to get
\[\begin{array}{rcl}
dX_t&=&\ds \eta'(\xi_t)d\xi_t+\frac{1}{2}\eta''(\xi_t)\,d\!<\xi,\xi>_t\\[3mm]
&=&\ds \frac{b(\xi_t)}{\sigma(\xi_t)}dt+dW_t-\frac{\sigma'(\xi_t)}{2}dt \\[3mm]
&=&\ds
\underbrace{\left(\frac{b(\eta^{-1}(X_t))}{\sigma(\eta^{-1}(X_t))}-\frac{\sigma'(\eta^{-1}(X_t))}{2}\right)}_{a(X_t)}dt+dW_t
\end{array}
\]
So $\xi_t=\eta^{-1}(X_t)$ where $(X_t)_t$ is a solution of the
stochastic differential equation
\begin{equation}
\left\{ \begin{array}{rcl}
dX_t&=&a(X_t)dt+dW_t\\
X_0&=&x.
\end{array}\right.
\label{eqn:SDE1}
\end{equation}
Thus, without loss of generality, one can start from
equation~(\ref{eqn:SDE1}) instead of~(\ref{eqn:SDEo}).

Let us denote by $(W_t^x)_{t\in[0,T]}$ the process
$(W_t+x)_{t\in[0,T]}$, by $\qq_{W^x}$ its law and by $\qq_X$ the law
of the process $(X_t)_{t\in[0,T]}$. From now on, we will denote by
$(Y_t)_{t\in
  [0,T]}$ the canonical process, that is the coordinate mapping on the set $C([0,T],\rr)$
of real continuous maps on $[0,T]$ (see~Revuz and Yor
\cite{RevuzYor} or~Karatzas and Shreve \cite{KaratzasShreve}).

One needs the following assumption to be true

\vspace{2mm}

\underline{\emph{Assumption 1}} : \textit{Under $\qq_{W^x}$, the
process}
\[L_t=\exp\left[\int_0^t a(Y_u)dY_u-\frac{1}{2}\int_0^t a^2(Y_u)du\right]\]
\textit{is a martingale.}

\vspace{4mm}

According to Rydberg \cite{Rydberg} (see the proof of
Proposition~\ref{prop:proof1} where we give his argument on a
specific example), a sufficient condition for this assumption to
hold is

-Existence and uniqueness in law of a solution to the
SDE~(\ref{eqn:SDE1}).

-$\ds \forall t \in [0,T], \int_0^t a^2(Y_u) du < \infty, \, \qq_X $
and $\qq_{W^x}$ almost surely on $C([0,T],\rr)$.

Thanks to this  assumption, one can apply the Girsanov theorem to
get that $\qq_X$ is absolutely continuous with respect to
$\qq_{W^x}$ and its Radon-Nikodym derivative is equal to
\[\frac{d\qq_X}{d\qq_{W^x}}=\exp\left[\int_0^Ta(Y_t)dY_t-\frac{1}{2}\int_0^Ta^2(Y_t)dt\right].\]

Consider $A$ the primitive of the drift $a$, and assume that

\vspace{2mm}

\underline{\emph{Assumption 2}} : \textit{$a$ is continuously
differentiable.}

\vspace{4mm}

Since, by Itô's lemma, $A(W^x_T)=A(x)+\int_0^T
a(W^x_t)dW^x_t+\frac{1}{2} \int_0^T a'(W^x_t)dt$, we have
\[\frac{d\qq_X}{d\qq_{W^x}}=\exp\left[A(Y_T)-A(x)-\frac{1}{2}\int_0^Ta^2(Y_t)+a'(Y_t)dt\right].\]

Before setting up an acceptance-rejection algorithm using this
Radon-Nikodym derivative, a last step is needed. To ensure the
existence of a density $h(u)$ proportional to
$\exp(A(u)-\frac{(u-x)^2}{2T})$, it is necessary and sufficient that
the following assumption holds

\vspace{2mm}

\underline{\emph{Assumption 3}} : \textit{The function $u \mapsto
\exp(A(u)-\frac{(u-x)^2}{2T})$ is integrable.}

\vspace{4mm}

Finally, let us define a process $Z_t$ distributed according to the
following law $\qq_Z$
\[\qq_Z= \int_\rr \mathcal{L}\Big((W^x_t)_{t \in [0,T]} | W^x_T=y\Big) h(y) dy\]

where the notation $\mathcal{L}(.|.)$ stands for the conditional
law. One has
\[\frac{d\qq_X}{d\qq_Z}=\frac{d\qq_X}{d\qq_{W^x}} \frac{d\qq_{W^x}}{d\qq_Z} =C \exp\left[-\frac{1}{2}\int_0^Ta^2(Y_t)+a'(Y_t)dt\right]\]

where $C$ is a normalizing constant. At this level, Beskos
\textit{et al.} \cite{BPR} need another assumption

\vspace{2mm}

\underline{\emph{Assumption 4}} : \textit{The function $\phi : x
\mapsto \frac{a^2(x)+a'(x)}{2}$ is bounded from below.}

\vspace{4mm}

Therefore, one can find a lower bound $k$ of this function and
eventually the Radon-Nikodym derivative of the change of measure
between $X$ and $Z$ takes the form
\begin{equation*}
\frac{d\qq_X}{d\qq_Z} = Ce^{-kT}
\exp\left[-\int_0^T\!\!\phi(Y_t)-k\,dt\right].
\end{equation*}

The idea behind the exact algorithm is the following : suppose that
one is able to simulate a continuous path $Z_t(\omega)$ distributed
according to $\qq_Z$ and let $M(\omega)$ be an upper bound of the
mapping $t \mapsto \phi(Z_t(\omega))-k$. Let $N$ be an independent
random variable which follows the Poisson distribution with
parameter $TM(\omega)$ and let $(U_i,V_i)_{i=1
  \dots N}$ be a sequence of independent random variables uniformly distributed on $[0,T] \times [0,M(\omega)]$. Then, the number of points  $(U_i,V_i)$ which fall
below the graph $\{(t,\phi(Z_t(\omega))-k); t \in [0,T]\}$ is equal
to zero with probability
$\exp\left[-\int_0^T\!\!\phi(Z_t(\omega))-k\,dt\right]$. Actually,
simulating the whole path $(Z_t)_{t\in[0,T]}$ is not necessary. It
is sufficient to determine an upper bound for $\phi(Z_t)-k$ since,
as pointed out by the authors, it is possible to simulate
recursively a Brownian motion on a bounded time interval by first
simulating its endpoint, then simulating its minimum or its maximum
and finally simulating the other points\footnote{In their paper, the
authors explain how to do such a decomposition of the Brownian
path.}. For this reason, one needs the following assumption for the
algorithm to be feasible :

\vspace{2mm}

\underline{\emph{Assumption 5}} : \textit{Either $\ds \limsup_{u \to
+\infty} \phi(u) < +\infty$ or $\ds \limsup_{u \to -\infty} \phi(u)
< +\infty$.}

\vspace{4mm}

Suppose for example that $\ds \limsup_{u \to +\infty} \phi(u) <
+\infty$. The exact algorithm of Bekos \textit{et al.} \cite{BPR}
then takes the following form :
\begin{algo}
$\,$
\begin{enumerate}
\item Draw the ending point $Z_T$ of the process $Z$ with respect to the density $h$.
\item Simulate the minimum $m$ of the process $Z$ given $Z_T$.
\item Fix an upper bound $M(m)=\sup\{\phi(u)-k;u \geq m\}$ for the mapping $t \mapsto \phi(Z_t)-k$.
\item Draw $N$ according to the Poisson distribution with parameter $TM(m)$
  and draw $(U_i,V_i)_{i=1 \dots N}$, a sequence of independent variables uniformly distributed on $[0,T] \times [0,M(m)]$.
\item Fill in the path of $Z$ at the remaining times $(U_i)_{i=1 \dots N}$.
\item Evaluate the number of points $(V_i)_{i=1\dots N}$ such that $V_i \leq \phi(Z_{U_i})-k$.\\
$\quad$ If it is equal to zero, then return the simulated path $Z$.\\
$\quad$ Else, return to step 1.
\end{enumerate}
\label{algo:EA}
\end{algo}
This algorithm gives exact skeletons of the process $X$, solution of
the SDE~(\ref{eqn:SDE1}). Once accepted, a path can be further
recursively simulated at additional times without any other
acceptance/rejection criteria. We also point out that the same
technique can be generalized by replacing the Brownian motion in the
law of the proposal $Z$ by any process that one is able to simulate
recursively by first simulating its ending point, its
minimum/maximum and then the other points. Also, the extension of
the algorithm to the inhomogeneous case, where the drift coefficient
$a$ in~(\ref{eqn:SDE1}), and therefore the function $\phi$, depend
on the time variable $t$, is straightforward given that the
assumptions presented above are appropriately modified.

\subsection{The unbiased estimator (U.E)\label{ece}}
In finance, the pricing of contingent claims often comes down to the
problem of computing an expectation of the form
\begin{equation}
  C_0=\ee\left(f(X_T)\right)\label{exp}
\end{equation}
where $X$ is a solution of the SDE~(\ref{eqn:SDE1}) and $f$ is a
scalar function such that $f(X_T)$ is square integrable. In a
simulation based approach, one is usually unable to exhibit an
explicit solution of this SDE and will therefore resort to numerical
discretization schemes, such as the Euler or Milstein schemes, which
introduce a bias. Of course, the exact algorithm presented above
avoids this bias. Here, we are going to present a technique which
permits to compute exactly the expectation~(\ref{exp}) while
assumptions 4 and 5 on the function $\frac{a^2+a'}{2}$ which appears
in the Radon-Nikodym derivative are relaxed.


Using the previous results and notations, we get, under the
assumptions 1 and 2, that
\begin{equation}
C_0=\ds \ee\left(f(W_T^x)
  \exp\left[A(W^x_T)-A(x)-\frac{1}{2}\int_0^Ta^2(W_t^x)+a'(W_t^x)dt\right]\right).
\label{etoile}
\end{equation}

In order to implement an importance sampling method, let us
introduce a positive density $\rho$ on the real line and a process
$(Z_t)_{t\in[0,T]}$ distributed according to the following law
$\qq_Z$
\begin{equation*}
  \qq_Z= \int_\rr \mathcal{L}\Big((W^x_t)_{t \in [0,T]} | W^x_T=y\Big) \rho(y)
  dy.
\end{equation*}

By~(\ref{etoile}), one has
\begin{equation}
C_0=\ds \ee\left(\psi(Z_T)
\exp\left[-\int_0^T\phi(Z_t)dt\right]\right) \label{FuncInt}
\end{equation}

where $\psi:z\mapsto f(z)
\frac{e^{A(z)-A(x)-\frac{(z-x)^2}{2T}}}{\sqrt{2\pi}\rho(z)}$ and
$\phi:z\mapsto \frac{a^2(z)+a'(z)}{2}$. We do not impose $\rho$ to
be equal to the density $h$ of the previous section. It is a free
parameter chosen in such a way that it reduces the variance of the
simulation.

In his first paper,~Wagner \cite{Wagner3_1} constructs an unbiased
estimator of the expectation~(\ref{FuncInt}) when $\psi$ is a
constant, $(Z_t)_{t\in[0,T]}$ is an $\rr^d-$valued Markov process
with known transition function and $\phi$ is a measurable function
such that $\ee\left(e^{\int_0^T |\phi(Z_t)|dt}\right) < +\infty$.
His main idea is to expand the exponential term in a power series,
then, using the transition function of the underlying Markov process
and symmetry arguments, he constructs a signed measure $\nu$ on the
space $\mathcal{Y}=\bigcup_{n=0}^{+\infty} ([0,T]\times\rr^d)^{n+1}$
such that the expectation at hand is equal to $\nu(\mathcal{Y})$.
Consequently, any probability measure $\mu$ on $Y$ that is
absolutely continuous with respect to $\nu$ gives rise to an
unbiased estimator $\zeta$ defined on $(\mathcal{Y},\mu)$ via
$\zeta(y)=\frac{d\nu}{d\mu}(y)$. In practice, a suitable way to
construct such an estimator is to use a Markov chain with an
absorbing state. Wagner also discusses variance reduction
techniques, specially importance sampling and a shift procedure
consisting on adding a constant $c$ to the integrand $\phi$ and then
multiplying by the factor $e^{-cT}$ in order to get the right
expectation. Wagner \cite{Wagner3_2} extends the class of unbiased
estimators by perturbing the integrand $\phi$ by a suitably chosen
function $\phi_0$ and then using mixed integration formulas
representation. Very recently, Beskos \textit{et al.} \cite{BPRF}
obtained a simplified unbiased estimator for~(\ref{FuncInt}), termed
Poisson estimator, using Wagner's idea of expanding the exponential
in a power series and his shift procedure. To be specific, the
Poisson estimator writes
\begin{equation}
  \psi(Z_T) e^{c_pT-cT} \prod_{i=1}^N \frac{c-\phi(Z_{V_i})}{c_P}
\end{equation}
where $N$ is a Poisson random variable with parameter $c_P$ and
$(V_i)_i$ is a sequence of independent random variables uniformly
distributed on $[0,T]$. Fearnhead \textit{et al.} \cite{FPR}
generalized this estimator allowing $c$ and $c_P$ to depend on $Z$
and $N$ to be distributed according to any positive probability
distribution on $\nn$. They termed the new estimator the generalized
Poisson estimator. We introduce a new degree of freedom by allowing
the sequence $(V_i)_i$ to be distributed according to any positive
density on $[0,T]$. This gives rise to the following unbiased
estimator for~(\ref{FuncInt}) :

\begin{lemma}
Let $p_Z$ and $q_Z$ denote respectively a positive probability
measure on $\nn$ and a positive probability density on $[0,T]$. Let
$N$ be distributed according to $p_Z$ and $(V_i)_{i \in \nn^*}$ be a
sequence of independent random variables identically distributed
according to the density $q_Z$, both independent from each other
conditionally on the process $(Z_t)_{t\in[0,T]}$. Let $c_Z$ be a
real number which may depend on $Z$. Assume that
\[\ee\left(|\psi(Z_T)| e^{-c_ZT} \exp\left[\int_0^T
  |c_Z-\phi(Z_t)|dt\right]\right) < \infty.\]

Then

\begin{equation}
\psi(Z_T) e^{-c_ZT}\frac{1}{p_Z(N)\, N!} \prod_{i=1}^N
\frac{c_Z-\phi(Z_{V_i})}{q_Z(V_i)} \label{eqn:method}
\end{equation}
is an unbiased estimator of $C_0$. \label{lemma1}
\end{lemma}

\begin{proof}
The result follows from

\begin{equation*}
\begin{array}{rcl}
\!\!\!\!\ds \ee\left(\!\!\psi(Z_T) e^{-c_ZT} \frac{1}{p_Z(N)\, N!}
\prod_{i=1}^N \frac{c_Z-\phi(Z_{V_i})}{q_Z(V_i)} \Big| (Z_t)_{t\in
[0,T]}\right)\!\!\!\!&\!\!=\!\!&\!\!\ds \psi(Z_T) e^{-c_ZT}
\sum_{n=0}^{+\infty} \frac{\left(\int_0^T
c_Z-\phi(Z_t)dt\right)^n}{p_Z(n) \, n!} \,
p_Z(n) \\[4mm]
&\!\!=\!\!&\!\!\ds\psi(Z_T)  \exp\left(-\int_0^T \phi(Z_t)dt\right).
\end{array}
\end{equation*}
\end{proof}

Using~(\ref{eqn:method}), one is now able to compute the expectation
at hand by a simple Monte Carlo simulation. The practical choice of
$p_Z$ and $q_Z$ conditionally on $Z$ is studied in the
appendix~\ref{PorQ}.

As pointed out in Fearnhead \textit{et al.} \cite{FPR}, this method
is an extension of the exact algorithm method since, under
assumptions 3, 4 and 5, the reinforced integrability assumption of
Lemma~\ref{lemma1} is always satisfied.

Indeed, suppose for example that $\ds \limsup_{u \to +\infty}
\phi(u) < +\infty$ and let $k$ be a lower bound of $\phi$, $m_Z$ be
the minimum of the process $Z$ and $M_Z$ an upper bound of
$\{\phi(u)-k, u \geq m_Z\}$. Then, taking $c_Z=M_Z+k$ in
Lemma~\ref{lemma1} ensures the integrability condition :
\[\begin{array}{rcl}
\ee\left(|\psi(Z_T)| e^{-(M_Z+k) T}
e^{\int_0^T|M_Z+k-\phi(Z_t)|dt}\right)&=&\ee\left(|\psi(Z_T)|
e^{-(M_Z+k) T}
e^{\int_0^T\!\!M_Z+k-\phi(Z_t)dt}\right)\\[5mm]
&=&\ee\left(|\psi(Z_T)| e^{-\int_0^T \phi(Z_t)dt}\right) < \infty
\end{array}\]
and hence, one is allowed to write that

\[C_0=\ee\left(\psi(Z_T) e^{-(M_Z+k) T} \frac{1}{p_Z(N) N!}\prod_{i=1}^N
  \frac{M_Z+k-\phi(Z_{V_i})}{q_Z(V_i)}\right).\]

Better still, the random variable $\psi(Z_T) e^{-(M_Z+k) T}
\frac{1}{p_Z(N) N!}\prod_{i=1}^N
  \frac{M_Z+k-\phi(Z_{V_i})}{q_Z(V_i)}$ is square integrable when $p_Z$ is the Poisson
  distribution with parameter $M_ZT+k$ and $q_Z$ is the uniform distribution on $[0,T]$
  since we have then
  \[\begin{array}{rcl}
    \ds \ee\left(\left(\psi(Z_T) e^{-(M_Z+k) T} \frac{1}{p_Z(N) N!}\prod_{i=1}^N
  \frac{M_Z+k-\phi(Z_{V_i})}{q_Z(V_i)}\right)^2\right)&=& \ds \ee\left(\psi^2(Z_T)
\prod_{i=1}^N \left(1- \frac{\phi(Z_{V_i})}{M_Z+k}\right)^2\right)\\[5mm]
&\leq& \ds \ee\left(\psi^2(Z_T)\right) < \infty.
\end{array}\]
The last inequality follows from the square integrability of $f$ :
whenever one is able to simulate from the density $h$, introduced in
the exact algorithm, by doing rejection sampling, there exists a
density $\rho$ such that $\psi$, which is equal to
$f(Z_T)\frac{h(Z_T)}{\rho(Z_T)}$ up to a constant factor, is
dominated by $f$ and so is square integrable.

The square integrability property is very important in that we use a
Monte Carlo method. We see that, whenever the exact algorithm is
feasible, the unbiased estimator of lemma~\ref{lemma1} is a
simulable square integrable random variable, at least for the
previous choice of $p_Z$ and $q_Z$.

\begin{remark}
  One can derive two estimators of $C_0$ from the result of Lemma~\ref{lemma1} :
  \[\begin{array}{rcl}
    \ds \delta_1&=&\ds \frac{1}{n} \sum_{i=1}^n f(Z_T^i) \frac{e^{A(Z_T^i)-A(x)-\frac{(Z_T^i-x)^2}{2T}}}{\sqrt{2\pi}\rho(Z_T^i)} e^{-c_ZT}\frac{1}{p_Z(N^i)\, N^i!}
\prod_{j=1}^{N^i} \frac{c_Z-\phi(Z^i_{V_j^i})}{q_Z(V_j^i)}\\[5mm]
\ds \delta_2&=&\ds \frac{\ds \sum_{i=1}^n f(Z_T^i)
  \frac{e^{A(Z_T^i)-A(x)-\frac{(Z_T^i-x)^2}{2T}}}{\sqrt{2\pi}\rho(Z_T^i)}  \frac{1}{p_Z(N^i)\, N^i!}
\prod_{j=1}^{N^i} \frac{c_Z-\phi(Z^i_{V_j^i})}{q_Z(V_j^i)}}{\ds
\sum_{i=1}^n
\frac{e^{A(Z_T^i)-A(x)-\frac{(Z_T^i-x)^2}{2T}}}{\sqrt{2\pi}\rho(Z_T^i)}
\frac{1}{p_Z(N^i)\, N^i!} \prod_{j=1}^{N^i}
\frac{c_Z-\phi(Z^i_{V_j^i})}{q_Z(V_j^i)}}.\end{array}\]

\end{remark}

\section{Application : the pricing of continuous Asian options}
In the \BS~model, the stock price is the solution of the following
SDE under the risk-neutral measure $\pp$
\begin{equation}
\frac{dS_t}{S_t}=(r-\delta)dt+\sigma dW_t \label{eqn:BS}
\end{equation}
where all the parameters are constant : $r$ is the short interest
rate, $\delta$ is the
dividend rate and $\sigma$ is the volatility.\\
Throughout, we denote $\gamma=r-\delta-\frac{\sigma^2}{2}$. The
path-wise unique solution of~(\ref{eqn:BS}) is
\[S_t=S_0 \, \exp\!\left(\sigma W_t
+\gamma t\right).\]

We consider an option with pay-off of the form
\begin{equation}
f\left(\alpha S_T + \beta \int_0^TS_tdt\right) \label{eqn:option}
\end{equation}
where $f$ is a given function such that $\ee\left(f^2\left(\alpha
S_T + \beta
    \int_0^TS_tdt\right)\right)<\infty$, $T$ is the maturity of the option and
$\alpha,\beta$ are two given non negative parameters\footnote{The
underlying of this option is a weighted
  average of the stock price at maturity and the running average of the stock price
  until maturity with respective weights $\alpha$ and $\beta T$.}. Note that for
$\alpha=0$, this is the pay-off of a standard continuous Asian
option.

The fundamental theorem of arbitrage-free pricing ensures that the
price of the option under consideration is
\[C_0=\ee \left(e^{-rT} f\left(\alpha S_T+\beta \int_0^T
S_u du\right)\right).\]

At first sight, the problem seems to involve two variables : the
stock price and the integral of the stock price with respect to
time. Dealing with the PDE associated with Asian option pricing,
Rogers and Rogers and Shi \cite{RogersShi} used a suitable change of
variables to reduce the spatial dimension of the problem to one. We
are going to use a similar idea.

Let
\begin{equation*}
\xi_t=\left(\alpha S_0+\beta S_0 \int_0^t e^{-\sigma W_u-\gamma u}
du\right)\, e^{\sigma W_t+\gamma t}.
\end{equation*}
We have that
\[\begin{array}{rcl}
\xi_t&=& \ds \alpha S_0 e^{\sigma W_t+\gamma t}+\beta S_0 \int_0^t
e^{\sigma (W_t-W_u)+\gamma (t-u)} du\\[2mm]
 &=& \ds \alpha S_0 e^{\sigma B_t+\gamma t}+\beta S_0 \int_0^t
e^{\sigma B_s+\gamma s} ds
\end{array}\]
where we set $B_s=W_t-W_{t-s}, \forall s \in [0,t]$. Clearly,
$(B_s)_{s\in [0,t]}$ is a Brownian motion and thus the following
lemma holds

\begin{lemma}
$\forall t \in [0,T], \, \xi_t$ and $\ds \alpha S_t+\beta \int_0^t
S_u du$ have the same law.
\end{lemma}

As a consequence
\begin{equation*}
C_0=\ee \left(e^{-rT} f(\xi_T)\right).
\end{equation*}
By applying Itô's lemma, we verify that the process $(\xi_t)_{t\geq
0}$ is a positive solution of the following 1-dimensional stochastic
differential equation for which path-wise uniqueness holds
\begin{equation}
\left\{\begin{array}{rcl}
d\xi_t&=&\beta S_0 dt + \xi_t (\sigma dW_t + (\gamma + \frac{\sigma^2}{2})dt)\\[2mm]
\xi_0&=&\alpha S_0.
\end{array}\right.
\label{eqn:XiSDE}
\end{equation}

We are thus able to value $C_0$ by Monte Carlo simulation without
resorting to discretization schemes using one of the exact
simulation techniques described in the previous section. In the case
$\alpha=0$, one has to deal with the fact that $\xi_t$ starts from
zero which is the reason why we distinguish two cases.

\subsection{The case $\alpha \neq 0$}

We are going to apply both the exact algorithm of Beskos \textit{et
al.} \cite{BPR} and the method based on the unbiased estimator of
lemma~\ref{lemma1}.

We make the following change of variables to have a diffusion
coefficient equal to 1 :

\begin{equation}
X_t=\frac{\log(\xi_t)}{\sigma} \, \Rightarrow \,
\left\{\begin{array}{rcl}
dX_t&=&(\frac{\gamma}{\sigma}+\frac{\beta S_0}{\sigma}e^{-\sigma X_t})dt+dW_t\\[2mm]
X_0&=&x \quad \text{ with } x=\frac{\log(\alpha S_0)}{\sigma}.
\end{array}\right.
\label{eqn:SDEy}
\end{equation}

Thus
\[C_0=\ee\left(e^{-rT}f(e^{\sigma X_T})\right).\]
The following proposition ensures that assumption 1 is satisfied.
\begin{proposition}
The process $(L_t)_{t\in [0,T]}$ defined by
\begin{equation*}
L_t=\exp\left[\int_0^T
  (\frac{\gamma}{\sigma}+\frac{\beta S_0}{\sigma}e^{-\sigma Y_t})
  \,dY_t-\frac{1}{2} \int_0^T (\frac{\gamma}{\sigma}+\frac{\beta
    S_0}{\sigma}e^{-\sigma Y_t})^2 dt\right]
\end{equation*}

is a martingale under $\qq_{W^x}$. \label{prop:proof1}
\end{proposition}

\begin{proof}
Under $\qq_{W^x}$, $(L_t)_{t\in[0,T]}$ is clearly a non-negative
local martingale and hence a super-martingale. Then, it is a true
martingale if and only if $\ee_{\qq_{W^x}}\left(L_T\right)=1$.

Checking the classical Novikov's or Kamazaki's criteria is not
straightforward. Instead, we are going to use the approach developed
by Rydberg \cite{Rydberg} (see also Wong and Heyde \cite{WongHeyde})
who takes advantage of the link between explosions of SDEs and the
martingale property of stochastic exponentials.

Let us define the following stopping times :
\begin{equation*}
\tau_n(Y)=\inf\left\{t \in \rr^+ \text{ such that }
  \int_0^t\left(\frac{\gamma}{\sigma}+\frac{\beta
    S_0}{\sigma}e^{-\sigma Y_u}\right)^2 du \geq
  n\right\},
\end{equation*}
with the convention $\inf\{\emptyset\}=+\infty$.

The stopped process $(L_{t\wedge \tau_n(Y)})_{t\in[0,T]}$ is a true
martingale under $\qq_{W^x}$ since Novikov's condition is fulfilled.
According to the Girsanov theorem, one can define a new probability
measure $\qq^n_X$, which is absolutely continuous with respect to
$\qq_{W^x}$, by its Radon-Nikodym derivative
\[\frac{d\qq^n_X}{d\qq_{W^x}}=L_{T \wedge \tau_n(Y)}.\]

Hence
\begin{equation*}
\ee_{\qq^n_X}\left(\mathbb{1}_{\{\tau_n(Y)>T\}}\right)=\ee_{\qq_{W^x}}\left(\mathbb{1}_{\{\tau_n(Y)>T\}}
L_{T \wedge \tau_n(Y)}\right).
\end{equation*}

Since $(\tau_n(Y))_{n\in \nn}$ is a non decreasing sequence, we can
pass to the limit in the right hand side
We get
\begin{equation*}
\lim_{n\to+\infty}
\qq_X^n\left(\tau_n(Y)>T\right)=\ee_{\qq_{W^x}}\left(\mathbb{1}_{\{\tau_\infty(Y)>T\}}
L_{T \wedge \tau_\infty(Y)} \right)
\end{equation*}

where $\ds \tau_\infty(Y)$ denotes the limit of the non decreasing
sequence $(\tau_n(Y))_{n\in \nn}$.

Under $\qq_{W^x}$, $(Y_t)_{t\in[0,T]}$ has the same law as a
Brownian motion starting from $x$ so $\tau_\infty(Y)=+\infty \,,
\qq_{W^x}$ almost surely, and consequently
\[\ee_{\qq_{W^x}}\big(L_T\big)=\lim_{n\to+\infty} \qq_X^n\left(\tau_n(Y)>T\right).\]

On the other hand, the Girsanov theorem implies that, under
$\qq^n_X$, $(Y_t)_{t\in[0,T \wedge \tau_n(Y)]}$ solves a SDE of the
form~(\ref{eqn:SDEy}). To conclude the proof, it is sufficient to
check that trajectorial uniqueness holds for this SDE. Indeed, the
law of $(Y_t)_{t\in[0,T\wedge \tau_n(Y)]}$ under $\qq^n_X$ is the
same as the law of $(Y_t)_{t\in[0,T\wedge
  \tau_n(Y)]}$ under $\qq_X$. Hence
\[\qq_X^n\left(\tau_n(Y)>T\right)=\qq_X\left(\tau_n(Y)>T\right) \,
\ov{\longrightarrow}_{n \to +\infty} \,
\qq_X\left(\tau_\infty(Y)>T\right).\]

Clearly, $\int_0^t\left(\frac{\gamma}{\sigma}+\frac{\beta
    S_0}{\sigma}e^{-\sigma Y_u}\right)^2 du < \infty, \, \qq_X$ almost surely, so
\[\ee_{\qq_{W^x}}\big(L_T\big)=\qq_X\left(\tau_\infty(Y)>T\right)=1\]

as required.

In order to check trajectorial uniqueness for the
SDE~(\ref{eqn:SDEy}), we consider two solutions $X^1$ and $X^2$. We
have that
\[
d(X^1_t-X^2_t)=\frac{\beta S_0}{\sigma}\left(e^{-\sigma
X_t^1}-e^{-\sigma
    X_t^2}\right) dt \Rightarrow d|X^1_t-X^2_t|=\frac{\beta S_0}{\sigma}
sign(X^1_t-X^2_t) \left(e^{-\sigma X_t^1}-e^{-\sigma
    X_t^2}\right) dt.\]

So
\[|X^1_t-X^2_t|=\frac{\beta S_0}{\sigma} \int_0^t sign(X^1_s-X^2_s) \left(e^{-\sigma X_t^1}-e^{-\sigma
    X_t^2}\right)ds \leq 0.\]
The last inequality follows from the fact that $x \mapsto e^{-\sigma
x}$ is a decreasing function. Finally, almost surely, $\forall t
\geq 0, \, X_t^1=X^2_t$ which leads to strong uniqueness.
\end{proof}

Consequently, thanks to the Girsanov theorem, we have
\begin{equation}
\frac{d\qq_X}{d\qq_{W^x}}=\exp\left[\int_0^T
  \underbrace{(\frac{\gamma}{\sigma}+\frac{\beta S_0}{\sigma}e^{-\sigma
      Y_t})}_{a(Y_t)} dY_t-\frac{1}{2} \int_0^T (\frac{\gamma}{\sigma}+\frac{\beta
    S_0}{\sigma}e^{-\sigma Y_t})^2 dt\right].
\label{RNder}
\end{equation}
Set $A(u)=\int_0^ua(x)dx=\frac{\gamma}{\sigma} u + \frac{\beta
S_0}{\sigma^2} (1-e^{-\sigma u})$. Then
\begin{equation*}
\frac{d\qq_X}{d\qq_{W^x}}=\exp\left[A(Y_T)-A(x)-\frac{1}{2}
\int_0^Ta^2(Y_t)+a'(Y_t) dt\right].
\end{equation*}
The function $u \mapsto
\exp\left(A(u)-\frac{(u-Y_0)^2}{2T}\right)=\exp\left(\frac{\gamma}{\sigma}
u + \frac{\beta S_0}{\sigma^2} (1-e^{-\sigma
u})-\frac{(u-Y_0)^2}{2T}\right)$ is clearly integrable so we can
define a new process $(Z_t)_{t\in [0,T]}$ distributed according to
the following law $\qq_Z$
\[\qq_Z= \int_\rr \mathcal{L}\Big((W_t)_{t \in
    [0,T]} | W_T=y\Big) h(y) dy\]
where the probability density $h$ is of the form
\begin{equation}
h(u)=C \exp\left(A(u)-\frac{(u-Y_0)^2}{2T}\right) \quad \text{ with
} C \text{ a
  normalizing constant.}
\label{eqn:TermDist}
\end{equation}

\begin{remark}
Simulating from this probability distribution is not difficult (see
the appendix~\ref{DistTerm} for an appropriate method of
acceptance/rejection sampling).
\end{remark}

We have
\begin{equation*}
\frac{d\qq_X}{d\qq_Z} =C \exp\left[-\int_0^T
\frac{1}{2}(a^2(Y_t)+a'(Y_t)) dt\right].
\end{equation*}
Set
$\phi(x)=\frac{a^2(x)+a'(x)}{2}=\frac{(\frac{\gamma}{\sigma}+\frac{\beta
    S_0}{\sigma}e^{-\sigma x})^2-\beta S_0 e^{-\sigma x}}{2}$. A direct calculation gives

\[\inf_{x \in \rr} \phi(x)=\left\{\begin{array}{ll}
\frac{\gamma^2}{2\sigma^2} & \text{ if } 2 \gamma \geq \sigma^2\\[2mm]
\phi\left(\frac{1}{\sigma} \log(\frac{2\beta S_0}{\sigma^2-2
\gamma})\right) & \text { otherwise.}
\end{array}\right.\]
Set $k=\inf_{x \in \rr} \phi(x)$. Finally, we get
\begin{equation*}
\frac{d\qq_X}{d\qq_Z}=Ce^{-kT} \exp\left[-\int_0^T \!\phi(Y_t)-k\,
dt\right]. \label{eqn:rejet_densite}
\end{equation*}
We check that
\[\begin{array}{l}
\ds \lim_{x \rightarrow +\infty} \phi(x)=\frac{\gamma^2}{2\sigma^2} \, < \infty\\[2mm]
\ds \lim_{x \to -\infty} \phi(x)=+\infty.
\end{array}\]

Hence we can apply the algorithm~\ref{algo:EA} to simulate exactly
$X_T$ and compute $C_0=\ee\left(e^{-rT}f(e^{\sigma X_T})\right)$ by
Monte Carlo. On the other hand, using~(\ref{RNder}) we get
\[C_0=\ee\left(e^{-rT}f(e^{\sigma W^x_T}) \exp\left[A(W^x_T)-A(x)-\frac{1}{2}\int_0^T
  a^2(W^x_t)+a'(W^x_t) dt\right]\right)\]
and we can also use the unbiased estimator presented in the previous
section to compute this expectation.
\begin{remark}
We also applied the exact algorithm based on a geometric Brownian
motion instead of the standard Brownian motion which seems more
intuitive given the form of the SDE~(\ref{eqn:XiSDE}). The algorithm
is feasible because we can simulate recursively a drifted Brownian
motion and therefore a geometric Brownian motion by an exponential
change of variables. The results we obtained were not different from
the first method.
\end{remark}

\subsubsection{Numerical computation}\label{sec:BS+Trap+KV}

\begin{figure}[!ht]
\begin{center}
\psfrag{TTTTTTTTTTTT}{$S_T$} \psfrag{t}{$\ds \alpha S_T+\beta
\int_0^T S_t dt$}
\psfrag{aaaaaaaaaaaaaaaaaaaaaaaaaaaaaaaaaaaaaaaaaaaaaaaaaaaaaaaaaaaaaaaaaaaa}{Exact
Simulation of the underlying : $\ds \alpha S_T+\beta \int_0^T S_t
dt$} \leavevmode \epsfig{file=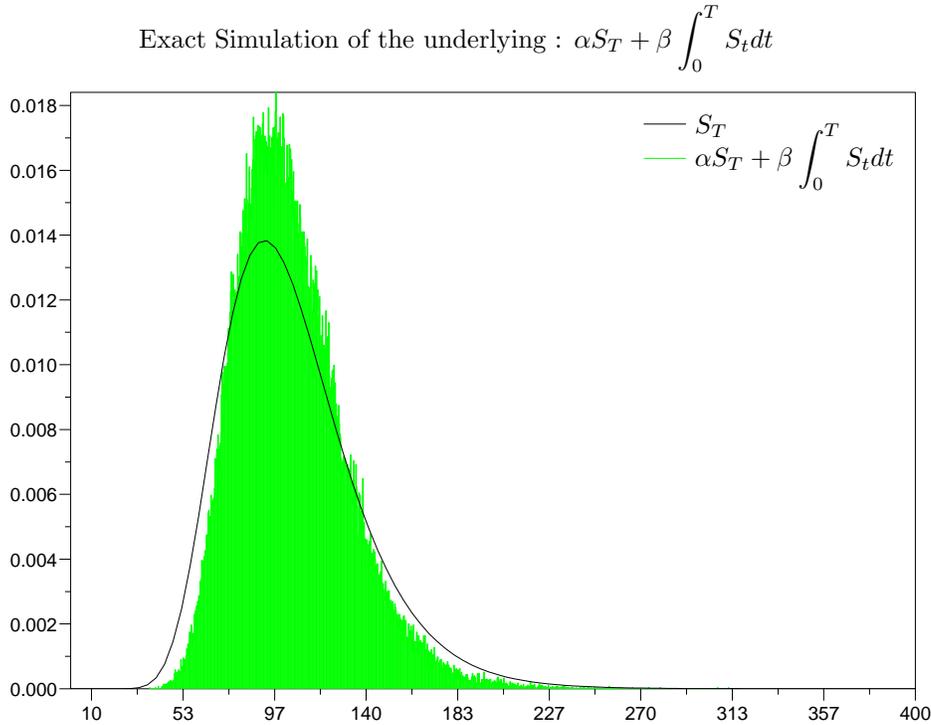,height=70ex}
\caption{Histogram of
  $10^5$ independent realizations of $\alpha S_T+\beta \int_0^T S_tdt$ for $\alpha=0.6$ and $\beta=0.4$ compared with
  the lognormal distribution of $S_T$.}

\label{fig:Distrib}
\end{center}
\end{figure}

For numerical tests, we consider the case
\[f(x)=(x-K)_+\]
which corresponds to the European call option with strike $K$. Using
the exact simulation algorithm presented above, we can simulate the
underlying $\alpha S_T+\beta \int_0^T S_t dt$ at maturity (see
Figure~\ref{fig:Distrib}). Then, all we have to do is a simple Monte
Carlo method to get the price of the option under consideration.
Using the unbiased estimator, we get

\[C_0=\ee\left(e^{-rT} (e^{\sigma Z_T}-K)_+
  \frac{e^{A(Z_T)-A(x)-\frac{(Z_T-x)^2}{2T}}}{\sqrt{2\pi}\rho(Z_T)} e^{-(M_Z+k) T} \frac{1}{p_Z(N) N!}\prod_{i=1}^N
  \frac{M_Z+k-\phi(Z_{V_i})}{q(V_i)} \right)\]
where $(Z_t)_{t\in[0,T]}, \rho, M_Z, k, p_Z$ and $q_Z$ are defined
as in section~\ref{ece}. In order to ensure square integrability, we
choose $p_Z$ to be a Poisson distribution with parameter $M_ZT+k$
and $q_Z$ to be the uniform distribution on $[0,T]$. For the density
$\rho$, a good choice is to consider the density that we use to
simulate from the distribution $h$ by rejection sampling.

We test these exact methods against a standard discretization scheme
with the variance reduction technique of Kemna and Vorst
\cite{KemnaVorst}. As pointed out by Lapeyre and Temam
\cite{LapeyreTemam}, the discretization of the integral by a simple
Riemannian sum is not efficient. Instead, we use the trapezoidal
discretization. In the sequel, we will denote this method by
Trap+KV. The table~\ref{tab:tab1e} gives the results we obtained for
the following arbitrary set of parameters : $S_0=100$, $K=100$,
$r=0.05$, $\sigma=0.3$, $\delta=0$, $T=1$, $\alpha=0.6$ and
$\beta=0.4$. The computation has been made on a computer with a 2.8
Ghz Intel Penthium 4 processor. We intentionally choose a large
number of simulations in order to show the influence of the number
of time steps when using a discretization scheme.

\begin{table}[!h]
\begin{center}
\begin{tabular}{|c|c|c|c|c|c|c|}
\hline Method & M & N & Acceptance rate & Price & C.I at $95\%$ & CPU \\
\hline
 & $10$ & & & $11.46$ & $[11.43,11.48]$ & 5 s\\ \cline{2-2} \cline{5-7}
Trap+KV & $20$ & $10^6$ & - & $11.46$ & $[11.43,11.49]$ & 9 s\\
\cline{2-2} \cline{5-7}
 & $50$ & & & $11.47$ & $[11.44,11.5]$ & 21 s\\
\hline
Exact Simulation & - & $10^6$ & 24\% & $11.46$ & $[11.43,11.5]$ & 81 s\\
\hline
U.E ($c_P=M_Z, c_Z=M_Z+k$) & - & $10^6$ & - & $11.46$ & $[11.43,11.49]$ & 17 s\\
\hline
U.E ($c_P=c_Z=1/T$)& - & $10^6$ & - & $11.46$ & $[11.43,11.49]$ & 6 s\\
\hline
\end{tabular}
\end{center}
\caption{Price of the option~(\ref{eqn:option}) using a standard
discretization technique and exact simulation methods.}
\label{tab:tab1e}
\end{table}

Empirical evidence shows that the exact simulation method is quite
slow. This is mainly due to the fact that the rejection algorithm
has a little acceptance rate ($24\%$ according to
table~\ref{tab:tab1e}). Using a geometric Brownian motion instead of
a standard Brownian motion did not improve the results. Also,
simulating recursively a Brownian path conditionally on its terminal
value and its minimum is time consuming.

The unbiased estimator is more efficient, especially when we can
avoid the recursive simulation of the Brownian path. To do so, we
choose for $p_Z$ a Poisson distribution with mean $c_PT$ where $c_P$
is a free parameter. If we assume that the integrability condition
in lemma~\ref{lemma1} holds, then we can write that
\[C_0=\ee\left(e^{-rT} (e^{\sigma Z_T}-K)_+
  \frac{e^{A(Z_T)-A(x)-\frac{(Z_T-x)^2}{2T}}}{\sqrt{2\pi}\rho(Z_T)} e^{c_PT-c_ZT}
  \prod_{i=1}^N \frac{c_Z-\phi(Z_{V_i})}{c_P} \right).\]

Regarding the dependence of the exact simulation method with respect
to the parameters $\alpha$ and $\beta$, it is intuitive that
whenever $\alpha >> \beta$, the method performs well since the
logarithm of the underlying is not far from the logarithm of the
geometric Brownian motion on which we do rejection-sampling. The
table~\ref{tab:tab2} confirms this intuition. We see that we cannot
apply the algorithm for small values of $\alpha$ and then let
$\alpha \to 0$ to treat the case $\alpha=0$.

\begin{table}[!h]
\begin{center}
\begin{tabular}{|c|c|c|c|c|c|}
\hline
\boldmath $\ds \frac{\alpha}{\alpha+\beta}$ &0.3 &0.4&0.5&0.6&0.7 \\
\hline
Acceptance Rate&0.003\% &0.47\%&5.66\%&24.43\%&53.85\% \\
\hline
\end{tabular}
\end{center}
\caption{Influence of the parameter $\frac{\alpha}{\alpha+\beta}$ on
the acceptance rate of the exact algorithm.} \label{tab:tab2}
\end{table}

\subsection{Standard Asian options : the case $\alpha = 0$ and $\beta > 0$}
A standard Asian option is a European option on the average of the
stock price over a determined period until maturity. An Asian call,
for example, has a pay-off of the form $(\frac{1}{T}\int_0^T S_u
du-K)_+$. With our previous notations, it corresponds to the case
$\alpha = 0$, $\beta=\frac{1}{T}$ and $f(x)=(x-K)_+$.

The change of variables we used above is no longer suitable because
it starts from zero when $\alpha=0$. Instead, we consider the
following new definition of the process $\xi$
\begin{equation}
\left\{\begin{array}{rcl}
\xi_t&=&\ds \frac{S_0}{t} \int_0^t e^{\sigma (W_t-W_u)+\gamma (t-u)} du\\[2mm]
\xi_0&=&\ds S_0.
\end{array}\right.
\label{eqn:Yfixed1}
\end{equation}
Obviously, the two variables $\xi_T$ and $\frac{1}{T} \int_0^T S_u
du$ have the same law. Hence, the price of the Asian option becomes
\begin{equation*}
C_0=\ee\left(e^{-rT} f\left(\frac{1}{T} \int_0^T S_u du
\right)\right)=\ee\left(e^{-rT} f(\xi_T)\right).
\end{equation*}

\begin{remark}
The pricing of floating strike Asian options is also straightforward
using this method. It is even more natural to consider these options
since it unveils the appropriate change of variables as we shall see
below.

Let us consider a floating strike Asian call for example. We have to
compute
\[C_0=\ee \left(e^{-rT} \big(\frac{1}{T} \int_0^T
S_u du -S_T\big)_+\right).\] Using $\widetilde{S}_t=S_t e^{\delta
t}$ as a numéraire (see the seminal paper of Geman \textit{et al.}
\cite{GemanKarouiRochet}), we immediately obtain that
\[C_0=\ee_{\pp_{\widetilde{S}}} \left( S_0 e^{-\delta T} \big(\frac{1}{T} \int_0^T \frac{S_u}{S_T} du -1\big)_+\right)\]
where $\pp_{\widetilde{S}}$ is the probability measure associated to
the numéraire $\widetilde{S}_t$. It is defined by its Radon-Nikodym
derivative $\frac{d
  \pp_{\widetilde{S}}}{d \pp}=e^{\sigma W_T- \frac{\sigma^2}{2}T}$.

Under $\pp_{\widetilde{S}}$, the process $B_t=W_t-\sigma t$ is a
Brownian motion and we can write that
\begin{equation*}
\begin{array}{rcl}
C_0&=&\ee_{\pp_{\widetilde{S}}} \left(S_0 e^{-\delta T} \big(\frac{1}{T} \int_0^T e^{\sigma(B_u-B_T)+(r-\delta+\frac{\sigma^2}{2})(u-T)} du -1\big)_+\right)\\[4mm]
&=&\ee \left(S_0 e^{-\delta T}\big(\frac{1}{T} \int_0^T
e^{\sigma(W_u-W_T)+(r-\delta+\frac{\sigma^2}{2})(u-T)} du -1\big)_+\right)\\[4mm]
&=&\ee \left(e^{-\delta T}\big(\xi_T-S_0\big)_+\right)
\end{array}
\end{equation*}
where $\xi_t$ is the process defined by~(\ref{eqn:Yfixed1}) but with
$\gamma=r-\delta+\frac{\sigma^2}{2}$. We see therefore that the
problem simplifies to the fixed strike Asian pricing problem.

\end{remark}

Let us write down the stochastic differential equation that rules
the process $(\xi_t)_{t\in [0,T]}$. Using Itô's lemma, we get
\begin{equation*}
\left\{\begin{array}{rcl}
d\xi_t&=&\frac{\xi_0-\xi_t}{t} dt+\xi_t \left(\sigma dW_t+(\gamma+\frac{\sigma^2}{2})dt\right)\\
\xi_0&=&S_0.
\end{array}\right.
\end{equation*}
Note that we are faced with a singularity problem near $0$ because
of the term $\frac{\xi_0-\xi_t}{t}$. We are going to reduce its
effect using another change of variables.

Using Itô's lemma, we show that
\begin{equation}
C_0=\ee\left(e^{-rT} f\left(S_0e^{X_T}\right)\right) \label{eqn:n}
\end{equation}
where $X_t=\log(\xi_t/\xi_0)$ solves the following SDE
\begin{equation}
\left\{\begin{array}{rcl}
dX_t&=& \sigma dW_t+\gamma dt+\frac{e^{-X_t}-1}{t}dt\\
X_0&=&0.
\end{array}\right.
\label{eqn:ZSDE}
\end{equation}
\begin{lemma}
Existence and strong uniqueness hold for the stochastic differential
equation~(\ref{eqn:ZSDE}).
\end{lemma}

\begin{proof}
Existence is obvious since we have a particular solution $X_t$. The
diffusion coefficient being constant and the drift coefficient being
a decreasing function in the spatial variable, we have also strong
uniqueness for the SDE (see the proof of
Proposition~\ref{prop:proof1}).
\end{proof}

Because of the singularity of the term $\frac{e^{-X_t}-1}{t}$ in the
drift coefficient, the law of $(X_t)_{t \geq 0}$ is not absolutely
continuous with respect to the law of $(\sigma W_t)_{t \geq 0}$.
That is why we now define $(Z_t)_{t \geq 0}$ by the following SDE
with an affine inhomogeneous drift coefficient :
\begin{equation}
\left\{\begin{array}{rcl}
dZ_t&=&\ds \sigma dW_t+\gamma dt-\frac{Z_t}{t}dt\\[2mm]
Z_0&=&X_0=0.
\end{array}\right.
\label{eqn:ZbarSDE}
\end{equation}
The drift coefficient exhibits the same behavior as the one
in~(\ref{eqn:ZSDE}) in the limit $t\to 0$ in order to ensure the
desired absolute continuity property. It is affine in the spatial
variable so that $(Z_t)_{t \geq 0}$ is a Gaussian process and as
such is easy to simulate recursively.
\begin{lemma}\label{lemma:order}
The process
\begin{equation}
{Z}_t=\frac{\sigma}{t}\int_0^ts dW_s+\frac{\gamma}{2}t
\label{eqn:Zbar}
\end{equation}
is the unique solution of the stochastic differential
equation~(\ref{eqn:ZbarSDE}).
\end{lemma}

\begin{proof}
Using Itô's Lemma, we easily check that $Z_t$ given
by~(\ref{eqn:Zbar}) is a solution of~(\ref{eqn:ZbarSDE}). Again,
constant diffusion coefficient and decreasing drift coefficient
ensures strong uniqueness.
\end{proof}

\begin{remark}\label{KVcv}
For the computation of the price
$C_0=\ee\left(e^{-rT}(S_0e^{X_T}-K)_+\right)$ of a standard Asian
call option, the random variable $e^{-rT}(S_0e^{Z_T}-K)_+$ provides
a natural control variate. Indeed, since $Z_T$ is a Gaussian random
variable with mean $\frac{\gamma}{2}T$ and variance $\frac{\sigma^2
T}{3}$, one has
\[\ee\left(e^{-rT}(S_0e^{Z_T}-K)_+\right)=S_0e^{(\frac{\gamma}{2}+\frac{\sigma^2}{6}-r)T}
\mathcal{N}\left(d+\sigma\sqrt{\frac{1}{3}T}\right)-Ke^{-rT}\mathcal{N}(d)\]
where $\mathcal{N}$ is the cumulative standard normal distribution
function and
$d=\frac{\log(S_0/K)+\frac{\gamma}{2}T}{\sigma\sqrt{\frac{1}{3}T}}$.

Notice that in~Kemna and Vorst \cite{KemnaVorst}, the authors suggest the use of the control variate\\
$e^{-rT}\left(S_0 \exp\left(\frac{1}{T} \int_0^T \sigma W_t + \gamma
t\,  dt\right)-K\right)_+$ which has the same law than
$e^{-rT}\left(S_0 e^{Z_T}-K\right)_+$ as \\$\ds \frac{1}{T} \int_0^T
\sigma W_t + \gamma t\, dt$ is also a Gaussian variable with mean
$\frac{\gamma}{2}T$ and variance $\frac{\sigma^2 T}{3}$.

\end{remark}

In order to define a new probability measure under which $(Z_t)_{t
\geq 0}$ solves the SDE~(\ref{eqn:ZSDE}), one introduces

\begin{equation*}
L_t=\exp\left[\int_0^t\frac{e^{-{Z}_s}-1+{Z}_s}{\sigma s}
  dW_s-\frac{1}{2} \int_0^t\left(\frac{e^{-{Z}_s}-1+{Z}_s}{\sigma
      s}\right)^2 ds\right].
\end{equation*}
Because of the singularity of the coefficients in the neighborhood
of $s=0$, one has to check that the integrals in $L_t$ are well
defined. This relies on the following lemma

\begin{lemma}
Let $\epsilon > 0$. In a random neighborhood of $s=0$, we have

\[|Z_s| \leq c s^{\frac{1}{2}-\epsilon} \text{ and } |X_s| \leq c
s^{\frac{1}{2}-\epsilon}\] where $c$ is a constant depending on
$\sigma$,$\gamma$ and $\epsilon$. \label{Z_X}
\end{lemma}
Since $\forall \epsilon > 0$,

\[\forall z \leq c s^{\frac{1}{2}-\epsilon},
\left(\frac{e^{-z}-1+z}{\sigma s}\right)^2 \leq C s^{-4 \epsilon},\]
we can choose $\epsilon<\frac{1}{4}$ to deduce that $L_t$ is well
defined.

\begin{proof}
We easily check that the Gaussian process $(B_t)_{t \in [0,T]}$
defined by $B_t=\ds \int_0^{(3t)^{\frac{1}{3}}} s dW_s$ is
  a standard Brownian motion. Thanks to the law of iterated logarithm for the Brownian
  motion (see for example~Karatzas and Shreve \cite{KaratzasShreve} p. 112), there exists $t_1(\omega)$ such that\footnote{$\omega$ is an element of the underlying probability space $\Omega$.},
\[\forall t \leq t_1(\omega), |B_t(\omega)| \leq t^{\frac{1}{2}-\frac{\epsilon}{3}}.\]
Therefore,
\[\forall t \leq (3t_1(\omega))^{\frac{1}{3}}, \quad |{Z}_t(\omega)| =\big|\frac{\sigma}{t}
B_{\frac{t^3}{3}}(\omega)+\frac{\gamma}{2}t\big| \leq
\frac{\sigma}{3^{\frac{1}{2}-\frac{\epsilon}{3}}}
t^{\frac{1}{2}-\epsilon}+\frac{\gamma}{2} t.\] Taking
$c=\max(\frac{\sigma}{3^{\frac{1}{2}-\frac{\epsilon}{3}}},\frac{\gamma}{2})$
yields
\[\forall t \leq (3t_1(\omega))^{\frac{1}{3}} \wedge 1, \quad |{Z}_t(\omega)|
  \leq c t^{\frac{1}{2}-\epsilon}.\]

On the other hand, recall that $ \ds
X_t=\log({\xi_t}/{\xi_0})=\log\left(\frac{1}{t}
  e^{\sigma W_t+\gamma t} \int_0^t e^{-\sigma W_u-\gamma u} du\right)$. So, using the
law of iterated logarithm for the Brownian motion, we deduce that
there exists $t_2(\omega)$ such that
\[\forall t \leq t_2(\omega), \quad 0 \leq \frac{1}{t}
  e^{\sigma W_t(\omega)+\gamma t} \int_0^t e^{-\sigma W_u(\omega)-\gamma u} du \leq \frac{1}{t}
  e^{\sigma t^{\frac{1}{2}-\epsilon}+\gamma t} \int_0^t e^{\sigma u^{\frac{1}{2}-\epsilon}-\gamma u} du.\]

Denote $g(t)=\frac{1}{t}
  e^{\sigma t^{\frac{1}{2}-\epsilon}+\gamma t} \int_0^t e^{\sigma
    u^{\frac{1}{2}-\epsilon}-\gamma u} du$ and let us investigate the order in
time near zero of this function. We have that
\[\begin{array}{rcl}
\ds e^{\sigma t^{\frac{1}{2}-\epsilon}+\gamma t}&=&\ds 1+\sigma t^{\frac{1}{2}-\epsilon}+\mathcal{O}(t^{1-2\epsilon})\\
\ds \int_0^t e^{\sigma u^{\frac{1}{2}-\epsilon}-\gamma u} du&=& \ds t+\frac{\sigma}{\frac{3}{2}-\epsilon} t^{\frac{3}{2}-\epsilon}+\mathcal{O}(t^{2-2\epsilon})\\
\end{array}\]
hence
\[g(t)=1+(\sigma+\frac{\sigma}{\frac{3}{2}-\epsilon})
t^{\frac{1}{2}-\epsilon}+\mathcal{O}(t^{1-2\epsilon}),\] so $\ds
X_t(\omega) \leq \log\left(g(t)\right) \ov{\sim}_{t \to 0}
(\sigma+\frac{\sigma}{\frac{3}{2}-\epsilon})
t^{\frac{1}{2}-\epsilon}$, which ends the proof for $X_t$.
\end{proof}

\begin{proposition}
$(L_t)_{t\in[0,T]}$ is a martingale and, consequently, for all $g :
\mathcal{C}([0,T]) \to \rr$ measurable, the random variables
$g((X_t)_{0 \leq t \leq T})$ and $g((Z_t)_{0 \leq t \leq T}) L_T$
are simultaneously integrable and then
\[\ee\Big(g((X_t)_{0 \leq t \leq T})\Big)=\ee\Big(g((Z_t)_{0 \leq t \leq T}) L_T\Big).\]
\label{prop:LAprop}
\end{proposition}

\begin{proof}
The proof is similar to the proof of Proposition~\ref{prop:proof1}.

We have already shown existence and strong uniqueness for both
SDE~(\ref{eqn:ZSDE}) and~(\ref{eqn:ZbarSDE}). Showing that the
stopping time
\[\tau_n(Y)=\inf\left\{t \in \rr^+ \text{ such that }
  \int_0^t\left(\frac{e^{-Y_s}-1+Y_s}{\sigma s}\right)^2 ds \geq
  n\right\}, \text{ with the convention }\inf\{\emptyset\}=+\infty,
\]
 have infinite limits when $n$ tends to $+\infty$, $\qq_X$ and $\qq_Z$ almost surely,
 follows from the previous lemma.

\end{proof}

One has
\begin{equation*}
L_T=\exp\left[\int_0^T\frac{e^{-{Z}_t}-1+{Z}_t}{\sigma^2 t}
  d{Z}_t-\int_0^T\frac{e^{-{Z}_t}-1+{Z}_t}{\sigma^2
      t} \left(\frac{e^{-{Z}_t}-1+{Z}_t}{2t}+\gamma-\frac{{Z}_t}{t}\right) dt\right].
\end{equation*}

Set $A(t,z)=\ds \frac{1-z+\frac{z^2}{2}-e^{-z}}{\sigma^2t}$. The
function $A:\,]0,T]\times \rr \rightarrow \rr$ is continuously
differentiable in time and twice continuously differentiable in
space. So, we can apply Itô's Lemma on the interval $[\epsilon,T]$
for $\epsilon > 0$ :
\[A(T,{Z}_T)=A(\epsilon,{Z}_\epsilon)+\int_\epsilon^T\frac{e^{-{Z}_t}-1+{Z}_t}{\sigma^2 t}
  d{Z}_t-\int_\epsilon^T\frac{1-{Z}_t+\frac{{Z}_t^2}{2}-e^{-{Z}_t}}{\sigma^2 t^2}dt+\int_\epsilon^T\frac{1-e^{-{Z}_t}}{2t}dt\]
Using the lemma~\ref{lemma:order}, we let $\epsilon \to 0$ to obtain
\[A(T,{Z}_T)=\int_0^T\frac{e^{-{Z}_t}-1+{Z}_t}{\sigma^2 t}
  d{Z}_t-\int_0^T\frac{1-{Z}_t+\frac{{Z}_t^2}{2}-e^{-{Z}_t}}{\sigma^2 t^2}dt+\int_0^T\frac{1-e^{-{Z}_t}}{2t}dt.\]

Then
\begin{equation*}
L_T=\exp\left[A(T,{Z}_T)-\int_0^T\phi(t,{Z}_t)dt\right]
\end{equation*}
where $\phi$ is the mapping
\begin{equation}
\phi(t,z)=\frac{e^{-z}-1+z-\frac{z^2}{2}}{\sigma^2t^2}+\frac{1-e^{-z}}{2t}+\frac{e^{-z}-1+z}{\sigma^2t}
\left(\frac{e^{-z}-1+z}{2t}+\gamma-\frac{z}{t}\right).\label{phi}
\end{equation}

By~(\ref{eqn:n}) and Proposition~\ref{prop:LAprop}, we get
\begin{equation}
\label{eq_Gob}
C_0=\ee\left(e^{-rT}f(S_0e^{{Z}_T})\exp\left[A(T,{Z}_T)-\int_0^T\phi(t,{Z}_t)dt\right]\right).
\end{equation}

Since for each $\ds t >0, \lim_{z \to -\infty} \phi(t,z) = +\infty$
and $\ds \lim_{z \to +\infty} \phi(t,z) = -\infty$, it is not
possible to apply the exact algorithm. One can use the unbiased
estimator, at least theoretically, if there exists a random variable
$c_Z$ measurable with respect to $Z$ such that
\[{\ee\left(e^{A(T,{Z}_T)-(r+c_Z)T} |f(S_0e^{{Z}_T})| e^{\int_0^T
|c_Z-\phi(t,Z_t)|dt}\right) <\infty}.\] Unfortunately, this
reinforced integrability condition is never satisfied :

\begin{lemma}
\label{lem_gob} Assume that $f$ is a non identically zero function.
Let $p_Z$ and $q_Z$ denote respectively a positive probability
measure on $\nn$ and a positive probability density on $[0,T]$. Let
$N$ be distributed according to $p_Z$ and $(U_i)_{i \in \nn^*}$ be a
sequence of independent random variables identically distributed
according to the density $q_Z$, both independent conditionally on
the process $(Z_t)_{t\in [0,T]}$. Then the random variable
\begin{equation}e^{A(T,{Z}_T)-rT} f(S_0e^{{Z}_T})  \, \frac{1}{p_Z(N) \, N!} \prod_{i=1}^N
  \frac{-\phi(U_i,{Z}_{U_i})}{q_Z(U_i)}\label{conjecture1}\end{equation}is non integrable.
\end{lemma}

\begin{proof}
By conditioning on $Z$, one has
\[\begin{array}{rcl}
\Delta:=\ee\left(\frac{e^{A(T,{Z}_T)-rT} |f(S_0e^{{Z}_T})|}{p_Z(N)
\, N!} \prod_{i=1}^N
  \frac{|\phi(U_i,{Z}_{U_i})|}{q_Z(U_i)}\right)&=&\ee\left(e^{A(T,{Z}_T)-rT} |f(S_0e^{{Z}_T})| e^{\int_0^T
|\phi(t,Z_t)|dt}\right)\\
&\geq& \ee\left(e^{A(T,{Z}_T)-rT} |f(S_0e^{{Z}_T})|
e^{\int_{\frac{T}{2}}^T |\phi(t,Z_t)|dt}\right)\end{array}\] One can
easily show that, $\forall z<0$ and $\forall t\in[\frac{T}{2},T],
\phi(t,z)\geq \overline{\phi}(z)$ where
\[\overline{\phi}(z)=\frac{e^{-z}-1+z-\frac{z^2}{2}}{\sigma^2(\frac{T}{2})^2}+\frac{e^{-z}-1+z}{\sigma^2\frac{T}{2}}
 \left(\frac{e^{-z}-1+z}{T}+\gamma^+-2\frac{z}{T}\right)
\]
Since $\ds
\overline{\phi}(z)\ov{\sim}_{-\infty}2\frac{e^{-2z}}{\sigma^2T^2}$,
there exists $c<0$ such that for all $z<c, \overline{\phi}(z)\geq
\frac{e^{-2z}}{\sigma^2T^2}$. Hence,
\[\begin{array}{rcl}
\Delta&\geq& \ee\left(e^{A(T,{Z}_T)-rT} |f(S_0e^{{Z}_T})|
e^{\frac{1}{\sigma^2T^2} \int_{\frac{T}{2}}^T
e^{-2Z_t}\ind_{\{Z_t<c\}}dt}\right)\\
&\geq& \ee\left(e^{A(T,{Z}_T)-rT} |f(S_0e^{{Z}_T})|
e^{-\frac{e^{-2c}}{2\sigma^2T}} e^{\frac{1}{\sigma^2T^2}
\int_{\frac{T}{2}}^T e^{-2Z_t}dt}\right)\end{array}\] Using Jensen's
inequality we get
\[\begin{array}{rcl}
\ds \Delta&\geq&\ds\ee\left(e^{A(T,{Z}_T)-rT}
|f(S_0e^{{Z}_T})|e^{-\frac{e^{-2c}}{2\sigma^2T}}
\exp\left({\frac{1}{2\sigma^2T}e^{-\frac{4}{T} \int_{\frac{T}{2}}^T Z_t dt}}\right)\right)\\[3mm]
\end{array}\]
We have seen in the proof of lemma~\ref{Z_X} that
$Z_t=\frac{\sigma}{t}B_{\frac{t^3}{3}}+\frac{\gamma}{2}t$ where
$(B_t)_{t\geq0}$ is a standard Brownian motion. So, conditionally on
$Z_T$, $\int_{\frac{T}{2}}^T Z_t dt$ is a gaussian random variable
and hence $\Delta=+\infty$.

\end{proof}

We are in a situation where $e^{A(T,{Z}_T)-rT} |f(S_0e^{{Z}_T})|\,
\ee\left[\left|\frac{1}{p_Z(N) \, N!} \prod_{i=1}^N
\frac{-\phi(U_i,{Z}_{U_i})}{q_Z(U_i)} \right|\Big|
(Z_t)_{t\in[0,T]}\right]$ is non integrable while $e^{A(T,{Z}_T)-rT}
|f(S_0e^{{Z}_T})|\, \left|\ee\left[\frac{1}{p_Z(N) \, N!}
\prod_{i=1}^N \frac{-\phi(U_i,{Z}_{U_i})}{q_Z(U_i)} \Big|
(Z_t)_{t\in[0,T]}\right]\right|$ is integrable since\\
$\ee\left(e^{-rT}|f(S_0e^{{Z}_T})|\exp\left[A(T,{Z}_T)-\int_0^T\phi(t,{Z}_t)dt\right]\right)<\infty$.
Then, a natural idea would consist in considering, for a given
$n\in\nn^*$, the random variable
\[e^{A(T,{Z}_T)-rT} |f(S_0e^{{Z}_T})|\,
\ee\left[\left|\frac{1}{n} \sum_{j=1}^n \frac{1}{p_Z(N_j) \, N_j!}
\prod_{i=1}^{N_j} \frac{-\phi(U_i^j,{Z}_{U_i^j})}{q_Z(U_i^j)}
\right|\Big| (Z_t)_{t\in[0,T]}\right]\] where $(N_j)_{1\leq j\leq
n}$ are independent variables having the same law as $N$ and
$\left((U^j_i)_{i \in \nn^*}\right)_{1\leq j\leq n}$ are independent
sequences having the same law as $\left(U_i\right)_{i \in \nn^*}$,
both independent conditionally on the process $(Z_t)_{t\in [0,T]}$.
The following general result tells us that this is not sufficient to
circumvent integrability problems.

\begin{lemma}
Let $Y$ and $Z$ be two real random variables and $g:\rr\to\rr$ a
given measurable function. Assume that $g(Z)\ee\left(Y|Z\right)$ is
integrable while $g(Z)\ee\left(|Y|\,|Z\right)$ is non integrable.
Then, when $(Y_i)_{1\leq i\leq n}$ is a sequence of independent
random variables having the same law as $Y$, $\forall n\in\nn^*$,
the random variable $g(Z)\ee\left(|\frac{1}{n}\sum_{i=1}^n
Y_i|\,|Z\right)$ is non integrable.
\end{lemma}

\begin{proof}
Denote by $e$, $e_1$ and $e_n$ three functions satisfying
\[\forall z\in\rr, \quad e(z)=\ee\left(Y |Z=z\right), \quad e_1(z)=\ee\left(|Y_1|
\,|Z=z\right) \quad \text{ and }
e_n(z)=\ee\left(\left|\frac{1}{n}\sum_{i=1}^n Y_i\right|
\,|Z=z\right)\] On the one hand, since $\int_\rr |g(z)| \, |e(z)|
\pp_Z(dz) <\infty$ and $\int_\rr |g(z)| \, e_1(z) \pp_Z(dz)
=+\infty$ , where $\pp_Z$ is the law of $Z$, we have that $\int_\rr
|g(z)| \, e_1(z)\ind_{\{e_1(z)\geq 2 |e(z)|\}} \pp_Z(dz)
=+\infty$.\\
On the other hand, $\forall z\in\rr$,
\[\begin{array}{rcl}
\ds e_n(z)&\geq&\ds \frac{1}{n}\left[\ee\left(\left|\sum_{i=1}^n
Y_i\right|\ind_{\{\forall 2\leq j\leq n, Y_j \geq 0\}}
|Z=z\right)+\ee\left(\left|\sum_{i=1}^n Y_i\right|\ind_{\{\forall
2\leq j\leq
n, Y_j <0\}} |Z=z\right)\right]\\[4mm]
&\geq&\frac{1}{n}\left[\ee\left(Y_1^+|Z=z\right)
\pp\left(Y_1\geq0|Z=z\right)^{n-1}+\ee\left(Y_1^-|Z=z\right)
\pp\left(Y_1<0|Z=z\right)^{n-1}\right]\\[4mm]
&=&\frac{1}{n}\left[\frac{e_1(z)+e(z)}{2}\pp\left(Y_1\geq0|Z=z\right)^{n-1}+\frac{e_1(z)-e(z)}{2}
\pp\left(Y_1<0|Z=z\right)^{n-1}\right]\\[4mm]
&\geq&\frac{1}{n}\left[\frac{e_1(z)}{4}\ind_{\{e_1(z)\geq2
|e(z)|\}}\pp\left(Y_1\geq0|Z=z\right)^{n-1}+\frac{e_1(z)}{4}
\ind_{\{e_1(z)\geq2
|e(z)|\}}\pp\left(Y_1<0|Z=z\right)^{n-1}\right]\\[4mm]
&\geq&\frac{e_1(z)}{n2^n}\ind_{\{e_1(z)\geq2 |e(z)|\}}
 \end{array}\]

Hence, $\ee\left[g(Z)\ee\left(\left|\frac{1}{n}\sum_{i=1}^n
Y_i\right|\,|Z\right)\right]=\int_\rr |g(z)| e_n(z)
\pp_Z(dz)=+\infty$.
\end{proof}

There is still hope yet. In the proof of Lemma \ref{lem_gob}, we saw
that integrability problems appear when $Z_t$ takes large negative
values so that $\phi(t,Z_t)$ tends rapidly towards $+\infty$. Since
$\ds \lim_{z \to +\infty} \phi(t,z) = -\infty$, one possible issue
is to split the function $\phi(t,Z_t)$ into a positive part and a
negative part. The first term can be handled by the exact simulation
technique whereas the second term, which as we shall see in the
following section presents no integrability problems, can be handled
by the unbiased estimator technique.

\subsubsection{An hybrid pseudo-exact method}
We rewrite (\ref{eq_Gob}) in the following form
\begin{equation}
\label{eq_Gob2}
C_0=\ee\left(e^{A(T,{Z}_T)-rT}f(S_0e^{{Z}_T})e^{\int_0^T\phi^-(t,{Z}_t)dt}
e^{-\int_0^T\phi^+(t,{Z}_t)dt}\right).
\end{equation}
Let $p_Z$ and $q_Z$ denote respectively a positive probability
measure on $\nn$ and a positive probability density on $[0,T]$. Let
$N$ be distributed according to $p_Z$ and $(U_i)_{i \in \nn^*}$ be a
sequence of independent random variables identically distributed
according to the density $q_Z$, both independent conditionally on
the process $(Z_t)_{t\in [0,T]}$. Note that, since
$e^{A(T,{Z}_T)-rT}f(S_0e^{{Z}_T})e^{\int_0^T|\phi^-(t,{Z}_t)|dt}
e^{-\int_0^T\phi^+(t,{Z}_t)dt}=e^{A(T,{Z}_T)-rT}f(S_0e^{{Z}_T})
e^{-\int_0^T\phi(t,{Z}_t)dt}$ is integrable, one has
\begin{equation}
C_0=\ee\left(e^{A(T,{Z}_T)-rT}f(S_0e^{{Z}_T})\,\frac{1}{p_Z(N)N!}\left(\prod_{i=1}^N
\frac{\phi^-(U_i,Z_{U_i})}{q_Z(U_i)}\right)\,
e^{-\int_0^T\phi^+(t,{Z}_t)dt} \right).
\end{equation}

\begin{remark}
There is no hope that this estimator is square integrable. Indeed,
one can show as in Lemma \ref{lem_gob} that
$\ee\left(e^{\int_0^T\left(\phi^-(t,Z_t)\right)^2dt}\right)=+\infty$
since $(\phi^-(t,z))^2$ is of order $z^4$ for large positive $z$.
\end{remark}

The idea then is to apply the exact simulation technique to simulate
an event with probability $e^{-\int_0^T\phi^+(t,{Z}_t)dt}$. Since
for each $t>0$, $\ds \lim_{z \to -\infty} \phi^+(t,z) = +\infty$,
one needs to bound from above $\phi^+(t,z)$, uniformly with respect
to $t\in[0,T]$, for $z>c$ where $c<0$ is a given constant. Thanks to
the following lemma, it is possible to do so but only uniformly with
respect to $t\in[\epsilon,T]$ for all $\epsilon>0$ :
\begin{lemma}
For all $0<t\leq T$,
\[\sup_{z\geq 0} \phi^+(t,z) \leq \frac{\gamma^2}{\sigma^2}+\frac{\gamma}{\sigma^2
t}+\frac{1}{t}\left(\frac{1}{2}-\frac{\gamma}{\sigma^2}\right)^+\]and\[\forall
c<0, \sup_{z\in[c,0]}\phi^+(t,z)\leq
\frac{e^{-c}-1+c}{\sigma^2t^2}(1+\gamma^+t)
+\frac{(e^{-c}-1)^2}{2\sigma^2t^2}-\frac{c^2}{\sigma^2t^2}.\]
\label{lem_Gob2}
\end{lemma}

\begin{proof}
Let $z>0$. It is useful to distinguish two cases according to the
sign of $\gamma$ :
\begin{enumerate}
\item \underline{$\gamma \geq0$}

We rewrite $\phi$ in the following form
\[\phi(t,z)=\frac{e^{-z}-1+z-\frac{z^2}{2}}{\sigma^2t^2}+\frac{1-e^{-z}}{t}\left(\frac{1}{2}-\frac{\gamma}{\sigma^2}\right)+
\frac{\gamma
z}{\sigma^2t}-\frac{z^2-(z\wedge1)^2}{2\sigma^2t^2}+\frac{(e^{-z}-1)^2-(z\wedge1)^2}{2\sigma^2t^2}
\]First note that $\frac{e^{-z}-1+z-\frac{z^2}{2}}{\sigma^2t^2}\leq0$,
$\frac{1-e^{-z}}{t}\left(\frac{1}{2}-\frac{\gamma}{\sigma^2}\right)\leq\frac{1}{t}\left(\frac{1}{2}-\frac{\gamma}{\sigma^2}\right)^+$
and $\frac{(e^{-z}-1)^2-(z\wedge1)^2}{2\sigma^2t^2}\leq0$. Moreover,
\[\begin{array}{rcl}
\ds \frac{\gamma
z}{\sigma^2t}-\frac{z^2-(z\wedge1)^2}{2\sigma^2t^2}&=&\ds
\frac{1}{\sigma^2}\left(\gamma
\frac{z}{t}-\frac{1}{2}\left(\frac{z}{t}\right)^2+\frac{(\frac{z}{t}\wedge\frac{1}{t})^2}{2}\right)\\[3mm]
&\leq&\ds \left\{\begin{array}{rl} \frac{\gamma}{\sigma^2t}&\text{ if }\gamma t \leq 1\\
\frac{\gamma^2}{\sigma^2}&\text{ otherwise}\\
\end{array}\right.\\
\end{array}
\]
Consequently, $\phi^+(t,z)\leq
\frac{\gamma^2}{\sigma^2}+\frac{\gamma}{\sigma^2
t}+\frac{1}{t}\left(\frac{1}{2}-\frac{\gamma}{\sigma^2}\right)^+$.

\item \underline{$\gamma \leq0$}

Now we rewrite $\phi$ in the following form
\[\phi(t,z)=\frac{e^{-z}-1+z-\frac{z^2}{2}}{\sigma^2t^2}+\gamma
\frac{e^{-z}-1+z}{\sigma^2t}+\frac{(e^{-z}-1)^2-z^2}{2\sigma^2t^2}+\frac{1-e^{-z}}{2t}
\]
It is then easy to show that $\phi^+(t,z)\leq\frac{1}{2t}$.
\end{enumerate}

Note that $\frac{1}{2t}\leq
\frac{\gamma^2}{\sigma^2}+\frac{\gamma}{\sigma^2
t}+\frac{1}{t}\left(\frac{1}{2}-\frac{\gamma}{\sigma^2}\right)^+$.
Hence, gathering the two cases yields
the first part of the lemma.

Let now $z\in[c,0]$ for a given negative constant $c$. We rewrite
$\phi$ in the following form
\[\phi(t,z)=\underbrace{\frac{e^{-z}-1+z}{\sigma^2t^2}(1+\gamma^+t)
+\frac{(e^{-z}-1)^2}{2\sigma^2t^2}-\frac{z^2}{\sigma^2t^2}}_{\geq0
\text{ for }
z<0}+\underbrace{\frac{1-e^{-z}}{2t}-\gamma^-\frac{e^{-z}-1+z}{\sigma^2t}}_{\leq
0 \text{ for }z<0}.
\]
Since $\ds
\partial_z\left[\frac{e^{-z}-1+z}{\sigma^2t^2}(1+\gamma^+t)
+\frac{(e^{-z}-1)^2}{2\sigma^2t^2}-\frac{z^2}{\sigma^2t^2}\right]=\frac{1-e^{-2
z}-2z+t\gamma^+(1-e^{-z})}{t^2 \sigma ^2}$ is negative for all
$z<0$, one has that
\[\sup_{z\in[c,0]}\phi^+(t,z)\leq \frac{e^{-c}-1+c}{\sigma^2t^2}(1+\gamma^+t)
+\frac{(e^{-c}-1)^2}{2\sigma^2t^2}-\frac{c^2}{\sigma^2t^2}.\]
\end{proof}

This lemma suggests to apply the exact algorithm on $[\epsilon,T]$
for a fixed positive threshold $\epsilon$. It remains to handle the
time interval $[0,\epsilon[$. Thanks to the following lemma, we that
$\phi^+(t,Z_t)$ can be approximately bounded from above for small
$t$, almost surely, by a function of $t$. The idea is then to extend
the exact simulation algorithm by simulating an inhomogeneous
Poisson process. Of course, this hybrid method is no longer exact
since the positive threshold for which the upper bound holds is
random.

\begin{lemma}
For all $\eta>0$, there exists a random neighborhood of $t=0$ such
that
\begin{equation}
\phi^+(t,Z_t)\leq\left(\frac{2
c^3}{3\sigma^2}+\frac{c}{2}\right)t^{-\frac{1}{2}-\eta}
\label{expansion}
\end{equation}where
$c=\max(\frac{\sigma}{3^{\frac{1}{2}-\frac{\eta}{3}}},\frac{\gamma}{2})$.
\label{equivalent}
\end{lemma}

\begin{proof}
We rewrite~(\ref{phi}) this way
\begin{equation*}
\phi(t,z)=\left(\frac{1-e^{-z}}{2}+\gamma
\frac{e^{-z}-1+z}{\sigma^2}\right)\frac{1}{t}-\left(\frac{1-z+\frac{z^2}{2}-e^{-z}-\frac{1}{2}(e^{-z}-1+z)(e^{-z}-1-z)}{\sigma^2}\right)\frac{1}{t^2}
\end{equation*}
and make the following Taylor expansions
\[\frac{1-z+\frac{z^2}{2}-e^{-z}-\frac{1}{2}(e^{-z}-1+z)(e^{-z}-1-z)}{\sigma^2}=\frac{2}{3\sigma^2}z^3+\mathcal{O}(z^4)\]

\[\text{and }\frac{1-e^{-z}}{2}+\gamma
\frac{e^{-z}-1+z}{\sigma^2}=\frac{1}{2}z+\mathcal{O}(z^2).\] On the
other hand, we have seen in the proof of lemma~\ref{Z_X} that there
exists a random neighborhood of zero such that $Z_t\leq c
t^{\frac{1}{2}-\eta}$ where
$c=\max(\frac{\sigma}{3^{\frac{1}{2}-\frac{\eta}{3}}},\frac{\gamma}{2})$.
We conclude that, in a random neighborhood of zero,
\[\phi^+(t,Z_t)\leq\left(\frac{2 c^3}{3\sigma^2}+\frac{c}{2}\right)t^{-\frac{1}{2}-\eta}.\]
\end{proof}

\subsubsection{Numerical computation}
For numerical computation,  we are going to use the following set of
parameters~: $S_0=100$, $K=100$, $\sigma=0.2$, $r=0.1$, $\delta=0$
and $T=1$. To fix the ideas, let us consider a call option. The
price $C_0$ writes as follows
\[
C_0=\ee\left(e^{A(T,{Z}_T)-rT}\left(S_0e^{{Z}_T}-K\right)^+\left(e^{c_p}\prod_{i=1}^N
\frac{\phi^-(U_i,{Z}_{U_i})}{c_p}\right)
e^{-\int_0^T\phi^+(t,{Z}_t)dt}\right).
\] where $N \sim\mathcal{P}(c_p)$ and $(U_i)_{i\geq1}$ is an
independent sequence of independent random variables uniformly
distributed in $[0,T]$. The parameter $c_p>0$ is set to one in the
following. We give a description of the hybrid method we implement :
\begin{algo}
$\,$

On the time interval $I_j:=[\frac{T}{2^{j+1}},\frac{T}{2^{j}}]$,
\begin{enumerate}
\item Simulate $Z_{\frac{T}{2^{j+1}}}, Z_{\frac{T}{2^{j}}}$ and a
lower bound $m_j$ for the minimum of $(Z_t)_{t\in I_j}$ (use the
fact that $Z_t=\frac{\sigma}{t}B_{\frac{t^3}{3}} +\frac{\gamma}{2}t$
where $(B_t)_{t\geq 0}$ is a standard Brownian motion).
\item Find $M^j>0$ such that $\forall t\in I_j,
\phi^+(t,Z_t)\leq M^j$ (use Lemma \ref{lem_Gob2}).
\item Simulate an homogeneous spatial Poisson process on the
rectangle $I_j \times [0,M^j]$ and accept (respectively reject) the
trajectory simulated if the number of points falling below the graph
$(\phi^+(t,Z_t))_{t\in I_j}$ is equal to (respectively different
from) zero.
\end{enumerate}
Carry on this acceptance rejection algorithm until reaching a time
interval $I_J$ for a chosen $J\in \nn^*$. On the remaining time
interval $[0,\frac{T}{2^{J+1}}],$ use the same acceptance/rejection
algorithm but with an inhomogeneous spatial Poisson process this
time (use Lemma \ref{equivalent}).
\end{algo}

In table~\ref{tab:WithWithout}, we give the price obtained by our
method for different values of the positive threshold
$\epsilon=\frac{T}{2^{J+1}}$. The number $M$ of Monte Carlo
simulations is equal to $10^5$ and the true price is equal to
$7.042$ (computed using a Monte Carlo method with a trapezoidal
scheme and a Kemna-Vorst control variate technique).

\begin{table}[!h]
\begin{center}
\begin{tabular}{|c|c|c|}
\hline & Price & CPU \\
\hline $\epsilon=\frac{T}{2^2}$& 6.9394 & 7s\\
\hline $\epsilon=\frac{T}{2^4}$& 6.9590 &10s\\
\hline $\epsilon=\frac{T}{2^6}$& 6.9703 &13s\\
\hline $\epsilon=\frac{T}{2^8}$& 6.9952 &17s\\
\hline $\epsilon=\frac{T}{2^{10}}$& 7.0423 &21s\\
\hline
\end{tabular}
\end{center}
\caption{Price of the Asian call using the hybrid-pseudo exact
method.} \label{tab:WithWithout}
\end{table}

Clearly, the method is not yet competitive regarding computation
time. Nevertheless, unlike the usual discretization methods, it is
not prone to discretization errors.

\section{Conclusion}

In this article, we have applied two original Monte Carlo methods
for pricing Asian like options which have the following pay-off :
$(\alpha S_T + \beta \int_0^TS_tdt-K)_+$. In the case $\alpha \neq
0$, we applied both the algorithm of Beskos \textit{et al.}
\cite{BPR} and a method based on the unbiased estimator of Wagner
\cite{Wagner3_1} and more recently the Poisson estimator of Beskos
\textit{et al.} \cite{BPRF} and the generalized Poisson estimator of
Fearnhead \textit{et al.} \cite{FPR}. The numerical results show
that the latter performs the best. The more interesting case
$\alpha=0$, which corresponds to usual continuously monitored Asian
options, can not be treated using neither the exact algorithm, nor
the method of exact computation of expectation but we investigate an
hybrid pseudo-exact method which combines the two techniques. More
generally, this hybrid method is an extension of the two exact
methods and can be applied in other situations.

From a practical point of view, the main contribution of these
techniques is to allow Monte Carlo pricing without resorting to
discretization schemes. Hence, we are no longer prone to the
discretization bias that we encounter in standard Monte Carlo
methods for pricing Asian like options. Even though these exact
methods are time consuming, they provide a good and reliable
benchmark.

\vspace{1cm}

\bibliographystyle{plain}
\addcontentsline{toc}{section}{Bibliography}
\bibliography{biblioArticle1}

\newpage

\section{Appendix}
\subsection{The practical choice of $p$ and $q$ in the U.E method
\label{PorQ}} The best choice for the probability law $p$ of $N$ and
the common density $q$ of the variables $(V_i)_{i \geq 1}$ is
obviously the one for which the variance of the simulation is
minimum. In a very general setting, it is difficult to tackle this
issue. In order to have a first idea, we are going to restrict
ourselves to the computation of $\ds \ee\left(\frac{1}{p(N)\, N!}
\prod_{i=1}^{N}\frac{g(V_i)}{q(V_i)}\right)$ where $g:[0,T] \to
\rr$.

\begin{lemma}
When $g$ is a measurable function on $[0,T]$ such that $\ds 0
<\!\int_0^T\!\! |g(t)| dt <+\infty$, the variance of $\ds
\frac{1}{p(N)\, N!} \prod_{i=1}^{N}\frac{g(V_i)}{q(V_i)}$ is minimal
for
\[q_{opt}(t)=\frac{|g(t)|}{\int_0^T|g(t)|dt}\, \mathbb{1}_{[0,T]}(t) \text{ and } \,
p_{opt}(n)=\frac{\left(\int_0^T|g(t)|dt\right)^{n}}{n!} \,
\exp\left( -\int_0^T|g(t)|dt\right).
\]
\end{lemma}

\begin{proof}
Minimizing the variance in~(\ref{eqn:method}) comes down to
minimizing the expectation of the square of $ \ds \frac{1}{p(N)\,
N!} \prod_{i=1}^N
\frac{g(V_i)}{q(V_i)}$.\\
Set
\begin{equation*}
F(p,q)=\ee\left(\frac{1}{(p(N)\, N!)^2} \prod_{i=1}^N
\frac{g^2(V_i)}{q^2(V_i)}\right)=\sum_{n=0}^{+\infty}
\frac{\left(\int_0^T\frac{g^2(t)}{q(t)} dt\right)^n}{p(n)\, (n!)^2}.
\end{equation*}
Using Cauchy-Schwartz inequality we obtain a lower bound for
$F(p,q)$
\[\begin{array}{rcl}
\ds F(p,q)=\sum_{n=0}^{+\infty}
\left(\frac{\left(\int_0^T\frac{g^2(t)}{q(t)}
    dt\right)^{\frac{n}{2}}}{p(n)\, n!}\right)^2 p(n)&\geq&\ds \left(\sum_{n=0}^{+\infty} \frac{\left(\int_0^T\frac{g^2(t)}{q(t)}
    dt\right)^{\frac{n}{2}}}{n!}\right)^2\\[5mm]
&=&\ds \left(\sum_{n=0}^{+\infty}
\frac{\left(\int_0^T\left(\frac{g(t)}{q(t)}\right)^2
      q(t) dt\right)^{\frac{n}{2}}}{n!}\right)^2\\[5mm]
&\geq &\ds \left(\sum_{n=0}^{+\infty}
  \frac{\left(\int_0^T|g(t)|dt\right)^{n}}{n!}\right)^2\\[5mm]
&=&\ds \exp\left(2\int_0^T|g(t)|dt\right).
\end{array}\]
We easily check that this lower bound is attained for $q_{opt}$ and
$p_{opt}$.

\end{proof}

The optimal probability distribution $p_{opt}$ is the Poisson law
with parameter$\ds \int_0^T|g(t)|dt$. This justifies our use of a
Poisson distribution for $p$.

\subsection{Simulation from the distribution $h$ given
by~(\ref{eqn:TermDist}) \label{DistTerm}} Recall that

\begin{equation*}
h(u) = C \exp\left(A(u)-\frac{(u-X_0)^2}{2T}\right)=C
\exp\left(\frac{\gamma}{\sigma} u + \frac{\beta S_0}{\sigma^2}
(1-e^{-\sigma u})-\frac{(u-X_0)^2}{2T}\right)
\end{equation*}

where $C$ is a normalizing constant.

The expansion of the exponential $e^{-\sigma u}$  at the first order
yields
\begin{equation*}
h(u) \approx C \exp\left(\frac{\gamma}{\sigma} u + \frac{\beta
S_0}{\sigma} u-\frac{(u-X_0)^2}{2T}\right) =C
\exp\left(-\frac{(u-(X_0+\frac{T(\gamma+\beta
S_0)}{\sigma}))^2}{2T}\right).
\end{equation*}

This suggests to do rejection sampling using the normal distribution
with mean $X_0+\frac{T(\gamma+\beta S_0)}{\sigma}$ and variance $T$
as prior. Unfortunately, for a standard set of parameters, this
method gives bad results. Even a second order expansion of
$e^{-\sigma u}$ which also modifies the variance does not work.

In order to get round this problem, we evaluate the mode $u^*$ of
$h$. We have
\begin{equation*}
h'(u^*)=C \left(\frac{\gamma}{\sigma}+ \frac{\beta
S_0}{\sigma}e^{-\sigma
  u^*}-\frac{u^*-X_0}{T}\right) \, \exp\left(\frac{\gamma}{\sigma} u^* + \frac{\beta S_0}{\sigma^2}
(1-e^{-\sigma u^*})-\frac{(u^*-X_0)^2}{2T}\right).
\end{equation*}

So, $h'(u^*)=0$ if and only if
\[
\frac{\gamma}{\sigma}+ \frac{\beta S_0}{\sigma}e^{-\sigma
u^*}-\frac{u^*-X_0}{T}=0
\]

which writes
\begin{equation*}
\sigma(u^*-X_0-\frac{\gamma}{\sigma}T)e^{\sigma(u^*-X_0-\frac{\gamma}{\sigma}T)}=T\beta
S_0 e^{-\sigma X_0-\gamma T}.
\end{equation*}

The function $x\mapsto xe^x$ is continuous and increasing on
$[0,+\infty[$ and so is its inverse which we denote by $W$. Since $T
\beta S_0 e^{-\sigma X_0-\gamma T} \geq 0$, we deduce that $h$ is
unimodal and that its mode satisfies

\[u^*=\frac{\gamma T + W\left(\beta S_0 T e^{-\gamma T-\sigma X_0}\right)+\sigma
  X_0}{\sigma}.\]

The function $W$ is the well-known Lambert function, also called the
Omega function. It is uniquely valued on $[0,+\infty[$ and there are
robust and fast numerical methods based on series expansion for
approximating this function (see for example Corless \textit{et al.}
\cite{Corless}).

Numerical tests showed that performing rejection sampling using a
Gaussian distribution with variance  $T$ and mean $u^*$ instead of
$X_0+\frac{T(\gamma+\beta
  S_0)}{\sigma}$ gives plain satisfaction. In table~\ref{tab:h_reject}, we see that for arbitrary choice of
the parameter $\frac{\alpha}{\alpha+\beta}$, the acceptance rate of
the algorithm is always high (of order $70\%$) and that the
computation time is low.

\begin{table}[!h]
\begin{center}
\begin{tabular}{|c|c|c|c|}
\hline
$\frac{\alpha}{\alpha+\beta}$ &Nb of simulations&Acceptance rate & Computation time\\
\hline 0.2&& 61\% &$3$s \\ \cline{1-1} \cline{3-4} 0.5&$10^6$&
68\%&$3$s\\ \cline{1-1} \cline{3-4}
0.8&& 80\%&$2$s \\
\hline
\end{tabular}
\end{center}
\caption{Acceptance rate of the rejection algorithm of simulating
from the distribution $h$ in~(\ref{eqn:TermDist}) with $S_0=100,
\sigma=0.3, T=2$ and $r=0.1$.} \label{tab:h_reject}
\end{table}

\end{document}